\documentclass[conference]{IEEEtran}
\IEEEoverridecommandlockouts
\usepackage{cite}
\usepackage{amsmath,amssymb,amsfonts}
\usepackage{graphicx}
\usepackage{textcomp}
\usepackage{xcolor}
\usepackage{booktabs}
\usepackage{color}
\usepackage{multirow}
\usepackage{algorithm}
\usepackage{graphicx}
\usepackage{amsmath}
\usepackage{amsthm}
\usepackage{booktabs}
\usepackage{multirow}
\usepackage{array}  
\usepackage{subfigure}
\usepackage{threeparttable}
\usepackage{newfloat}
\usepackage{listings}
\usepackage{marvosym}
\usepackage{algpseudocode}
\usepackage{algorithmicx}
\usepackage{algorithm}
\usepackage{balance}

\pdfoutput=1

\def\BibTeX{{\rm B\kern-.05em{\sc i\kern-.025em b}\kern-.08em
		T\kern-.1667em\lower.7ex\hbox{E}\kern-.125emX}}
\begin{document}
	\pdfoutput=1
	
	\title{Edge-Enhanced Global Disentangled Graph Neural Network for Sequential Recommendation}
	\author{\IEEEauthorblockN{Yunyi Li\IEEEauthorrefmark{1},
			Pengpeng Zhao\IEEEauthorrefmark{1}, Guanfeng Liu\IEEEauthorrefmark{3}, Yanchi Liu\IEEEauthorrefmark{4}, Victor S. Sheng\IEEEauthorrefmark{2}, Jiajie Xu\IEEEauthorrefmark{1}, Xiaofang Zhou\IEEEauthorrefmark{6}
		}
		\IEEEauthorblockA{\IEEEauthorrefmark{1}School of Computer Science and Technology, Soochow University, Suzhou, China}
		\IEEEauthorblockA{\IEEEauthorrefmark{3}, Macquarie University, Sydney, Australia,\IEEEauthorrefmark{4}Rutgers University, New Jersey, USA}
		\IEEEauthorblockA{\IEEEauthorrefmark{2}Department of Computer Science, Texas Tech University, Lubbock, USA}
		\IEEEauthorblockA{\IEEEauthorrefmark{6}The Hong Kong University of Science and Technology, Hong Kong SAR, China}
		\IEEEauthorblockA{\IEEEauthorrefmark{1}yyliyyli@stu.suda.edu.cn, \{ppzhao,xujj\}@suda.edu.cn}
		{\IEEEauthorrefmark{3}guanfeng.liu@mq.edu.au}
		{\IEEEauthorrefmark{4}yanchi.liu@rutgers.edu}   {\IEEEauthorrefmark{2}victor.sheng@ttu.edu}      {\IEEEauthorrefmark{6}zxf@cse.ust.hk}
	}

	\maketitle
	
	\begin{abstract}
		Sequential recommendation has been a widely popular topic of recommender systems. Existing works have contributed to enhancing the prediction ability of sequential recommendation systems based on various methods, such as recurrent networks and self-attention mechanisms. However, they fail to discover and distinguish various relationships between items, which could be underlying factors which motivate user behaviors. In this paper, we propose an Edge-Enhanced Global Disentangled Graph Neural Network (EGD-GNN) model to capture the relation information between items for global item representation and local user intention learning. At the global level, we build a global-link graph over all sequences to model item relationships. Then a channel-aware disentangled learning layer is designed to decompose edge information into different channels, which can be aggregated to represent the target item from its neighbors. At the local level, we apply a variational auto-encoder framework to learn user intention over the current sequence. We evaluate our proposed method on three real-world datasets. Experimental results show that our model can get a crucial improvement over state-of-the-art baselines and is able to distinguish item features.
		
		
	\end{abstract}
	
	
	\section{Introduction}
	
	Recommender systems play a critical role in the fast-developing Internet age, aiming to predict the most likely items which users may be interested in. Collaborative filtering is an efficient and widely used approach in recommendation, which commits to capturing latent user and item features from historical interactions. Early works like Matrix Factorization (MF)~\cite{koren2009matrix} decompose a rating matrix into user and item embeddings to capture implicit semantics. As the scale of users and items increases rapidly in recent years, more deep learning models are proposed based on collaborative filtering to characterize plentiful user tastes over a large amount of items. For example, ~\cite{berg2017graph} and~\cite{wang2019neural} build user-item graphs to integrate the multi-hop relationship of interactions.~\cite{liang2018variational} and~\cite{li2017collaborative} introduce a variational auto-encoder framework into the model and infer the representation as a Gaussian distribution. 
	
	Sequential recommendation is an important part among recommender systems. It models user behaviors as a sequence of items instead of a set of items. Markov Chain (MC)~\cite{cheng2013you} is a classic method, which models short-term item transitions and predicts the next item a user may like. Factorized personalized Markov chain (FPMC)~\cite{rendle2010factorizing} combines the markov chain and the traditional matrix factorization together to model user preferences. With the development of deep learning networks, Recurrent Neural Networks (RNNs) have achieved successes in sequential recommendation. For example, Long Short-Term Memory (LSTM)~\cite{zhu2017next} is a common variation of RNN to enhance model’s ability of maintaining sequential information by memory cells. GRU4Rec~\cite{chen2018sequential} applies Gated Recurrent Units (GRU) to session-based recommendation by introducing session-parallel mini-batches. RNN-based methods face the challenge of maintaining long-range information. Then self-attention network is applied to sequential recommendations recently to capture both long-term and short-term dependencies. SASRec and BERT4Rec both get good prediction results with attention mechanisms. SASRec~\cite{kang2018self} is able to capture long-term dependencies because it takes into account the influential weights of a whole historical sequence. BERT4Rec~\cite{sun2019bert4rec} employs a deep bidirectional self-attention network with Cloze tasks to increase the efficiency of a transformer model.
	
\begin{figure}[t]
		\includegraphics[width=0.47\textwidth]{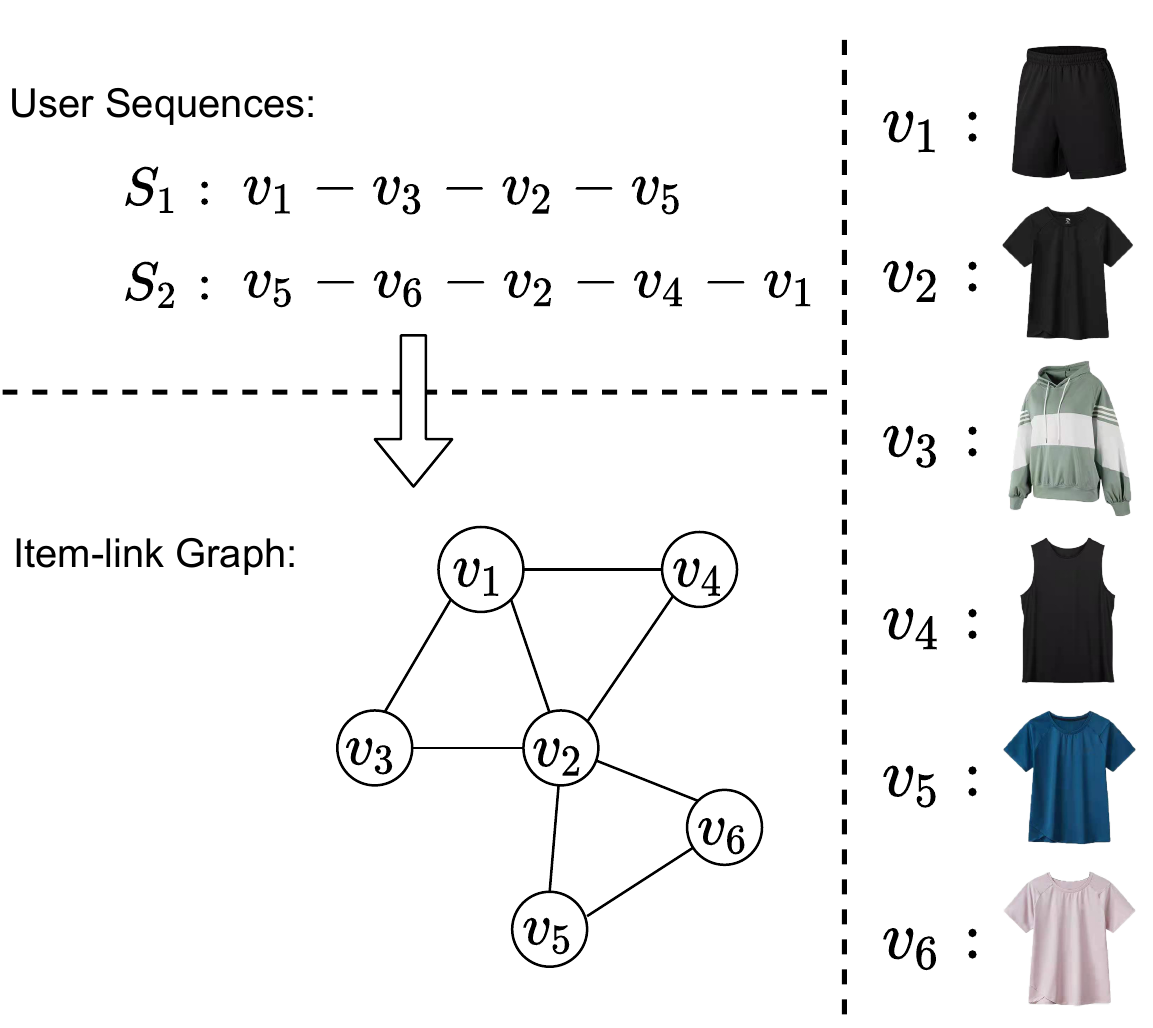}
		\caption{An example of item relationship patterns.} \label{intro}
	\end{figure}

	These previous works model user intentions with historical sequential interactions, ignoring dynamic underlying relationships behind items. The edges that link pairwise items contain abundant semantic information of factors why and how users choose one item after another. These underlying factors are related to real-world concepts, and one certain factor often plays a leading role in a single situation. For example, suppose there are two users interacting with six items, as shown in Figure~\ref{intro}. The link graph shows that item 2 is adjacent to all the other five items. But these edges are intuitively motivated by different factors. Item 2 is linked to items 1 and 4 because they are in the same color, while it is linked to 5 and 6 because they have short sleeves. Item 3 is connected to item 2 because it can be used as a T-shirt jacket. These different factors show the intention transformation of user behaviors, and also reveal the shared features of pairwise items. Therefore, recognizing and distinguishing the underlying item-link factors is able to enhance the expression ability of models, and disentangled representation learning~\cite{locatello2019challenging} is a common method proposed to achieve this goal.
	
	Disentangled representation learning has been of great popularity in many fields such as Computer Vision~\cite{hsieh2018learning, mu2021disentangled}, and it has been applied to recommender systems recently. The general purpose of disentangled representation learning is to separate the distinct and informative factors from the variations of data, where each unit is related to a single concept in the real world. A single change of one factor will lead to a change of the relevant unit. Many models have been proved to have the ability to learn disentangled representation and have been applied to fulfill realistic tasks. For example, by learning disentangled representation of a face image, we can obtain independent representations of different features of the face. Then we are able to identify whether a person in a picture has bangs, is wearing glasses, is smiling, and so on. Further, we can change these features directionally by modifying the values of the corresponding dimensions of these features. Therefore, learning disentangled representation can enhance the interpretability and controllability of model.
	
	The most prominent networks to learn disentangled representation are $\beta$-VAE and InfoGAN. $\beta$-VAE~\cite{higgins2016beta} adds a coefficient hyperparameter $\beta$ to the KL-divergence term in the objective of variational auto-encoder to encourage more factorized latent representations. This extra hyperparameter puts a heavy pressure on the posterior distribution to match the factorized prior distribution~\cite{burgess2018understanding}. InfoGAN~\cite{chen2016infogan} maximizes the mutual information between a fixed small subset of the GAN’s noise variables and observations.  With regard to recommender systems, Macrid-VAE~\cite{ma2019learning} infers high-level concepts of user intentions at the macro level and applies VAE to enhance disentanglement at the micro level. The authors also propose a self-supervised seq2seq training strategy for sequential recommendation~\cite{ma2020disentangled}, which compares user intentions between sub-sequences generated by an intention disentangled encoder. DGCF~\cite{wang2020disentangled} devises intent-aware interaction graphs to distinguish user intentions over different items, focusing on user-item relationships. However, these studies do not consider item-link relationship patterns, and fail to distinguish different user intentions behind sequences. As a result, the sequential model will be sensitive to noisy data and hardly interpretable.
	
	In this paper, we propose an \textbf{E}dge-Enhanced \textbf{G}lobal \textbf{D}isentangled \textbf{G}raph \textbf{N}eural \textbf{N}etwork (EGD-GNN) model to capture the item-link information. We model item representations and user intentions from both the global and the local levels. At the global level, we build a global item-link graph over all sequences, and each item-pair in the sequences is denoted as an edge in the graph. Figure~\ref{intro} shows an example of the construction of a global graph over two sequences. We apply the channel-aware mechanism to decompose edges into several channels, where each channel is correspond to an influential factor. The channels extract features specifically to one of the disentangled factors from the neighbors and aggregate different factors jointly to the target item. At the local level, we model disentangled user-intention representation over the current sequence. We first infer the latent variation as a Gaussian distribution in order to enforce disentanglement from the statistical perspective of variational auto-encoder. Then we use the channel-aware mechanism and aggregate item information through the edge channels from former items in the sequence. The aggregated item representation is used to express current user intention. We conduct experiments based on our proposed method on three real-world datasets and compare the predicting results with the state-of-art baselines. Results show that the EGD-GNN model not only outperforms the previous works in prediction tasks, but also forces the system to find good disentangled representations.
	
	The major contributions of this paper are summarized as follows:
	
	\begin{itemize}
		\item To the best of our knowledge, it is the first work to explore disentangled representation over the global and the local levels to learn factorized underlying factors of item relationships.
		
		\item We propose a disentangled graph neural network to infer the factors behind pairwise items which motivate user intentions. We apply the GNN model to the global graph and the local sequences to learn item link patterns, and employ variational auto-encoder taking advantage of its statistical property.
		
		\item We evaluate our model on three real-world datasets, and experimental results show our proposed model is able to achieve a good disentangled representation over sequences to indicate user intentions.
	\end{itemize}
	
	The rest of this paper is organized as follows. Firstly, we review related works corresponding to our work in Section~\ref{related}. Then we propose the problem formulation and definitions of our model and introduce preliminaries in Section~\ref{preliminary}. Next, we present the details of our proposed model in Section~\ref{methodology}. Section~\ref{experiment} records and visualizes experimental results, proving the effectiveness of our model. Finally, we conclude the paper and put forward a vision for future work in Section~\ref{conclusion}.

	\section{Related Work}\label{related}
	
	In this section, we present the recent works related to our model, including Sequential Recommendation, Disentangled Representation Learning, and Graph Neural Network.
	
	\subsection{Sequential Recommendation}
	
	Recommendation systems have been extensively popular over the last two decades. They aim to predict users’ preferences over historical behaviors. Matrix Factorization (MF)~\cite{koren2009matrix} is the most common framework for prediction, which learns user/item embeddings respectively to model latent relationships between users and items. Further study like SVD++~\cite{koren2008factorization} combines domain model and hidden factor model, and proposes a new globally optimized neighborhood model. Sequential recommendation is an important branch of recommendation systems. Given a chronological item sequence of user’s historical behaviors, sequential recommendation can predict the next item with which a user is likely to interact.
	
	Markov Chain (MC)~\cite{cheng2013you} is a classical model to capture short-term item transitions. FPMC~\cite{rendle2010factorizing} further combines Matrix Factorization and Markov Chain together to model both long-tern preferences and short-tern transitions. Fossil~\cite{he2016fusing} combines similarity-based models with high-order Markov Chains. TransRec~\cite{he2017translation} turns a user embedding into a translation vector and considers the three-order relationships between users, candidate items, and previous behaviors. With the proposal of Recurrent Neural Network (RNN), researchers have proposed redundant works based on this sequential framework and its variants. For example, Time-LSTM~\cite{zhu2017next} uses Long Short-Term Memory (LSTM) to model time intervals with time gates, and GRU4Rec~\cite{chen2018sequential} uses Gated Recurrent Units (GRU) to model click sequences for session-based recommendation. Recently, more deep neural networks have been applied to model sequence patterns. Caser~\cite{tang2018personalized} employs a Convolutional Neural Network (CNN) to capture sequential patterns as local features of images by embedding recent sequential items into the images. SASRec~\cite{kang2018self} finds the relevance between items adaptively using the self-attention mechanism. BERT4Rec~\cite{sun2019bert4rec} employs a deep bidirectional self-attention network with Cloze task to increase the efficiency of transformer model. SVAE~\cite{sachdeva2019sequential} leverages Variational Auto-encoder (VAE) to handle temporal information of sequences. However, these previous works do not distinguish the various contributions of neighbors over different aspects.

	\subsection{Disentangled Representation Learning}
	
	The purpose of learning disentangled representation is to find independent factors in the latent space. Each dimension of the representation has a specific and irrelevant meaning and is human-understandable. For example, learning disentangled representation over face pictures can get representations regarding eyes, hair, smiles, etc., while learning disentangled representation over landscape pictures can get representations regarding trees, sky, buildings, etc. Distinguishing such features from the representations can bring enhanced robustness, interpretability, and controllability. Therefore, it has been a popular task in many fields such as computer vision~\cite{hsieh2018learning} and topic modeling~\cite{lin2020graph}. During recent years, many methods have been proposed to improve the disentanglement learning ability. InfoGAN~\cite{chen2016infogan} realizes unsupervised learning of disentangled representation by introducing mutual information to constrain the latent variables. $\beta$-VAE~\cite{higgins2016beta} turns the perspective to information bottleneck and focuses on the KL-divergence in the VAE objective. Further studies like $\beta$-TCVAE~\cite{chen2018isolating} and FactorVAE~\cite{kim2018disentangling} decompose the KL-divergence and directly encourage factorized distribution by putting penalty on the total correlation.
	
	Recently, some studies turn attention to disentangled learning in recommendation. For instance, Macrid-VAE~\cite{ma2019learning} is the first work to learn disentangled representation from user-item interactions in recommender systems. At the macro level, it divides user intentions into several high-level concepts and categorizes each item into a concept. At the micro-level, it applies VAE framework to the encoder layer to encourage dimension independence. DGCF~\cite{wang2020disentangled} devises a disentangled graph model to learn user intents based on neural graph collaborative filtering. DICE~\cite{zheng2021disentangling} disentangles the interest and conformity representation with causal embedding. Ma et.al.~\cite{ma2020disentangled} performs self-supervision in the latent space to classify user intentions. It reconstructs future sequences as a whole using the sequence-to-sequence training strategy, instead of individual items in the future sequences. However, these works fail to maintain the item-link relationships and disentangle their influential factors. In this paper, we will solve this problem by introducing graph neural network into sequential recommendation.
	
	\subsection{Graph Neural Network}
	
	Graph Neural Network (GNN) is a classical learning network used to capture information of graph structure data. It has achieved great success in various tasks, such as node classification~\cite{kipf2016semi} and link prediction~\cite{taskar2003link}. Early work like ChebNet~\cite{defferrard2016convolutional} realizes fast localized convolutional filters on graphs by CNN and avoids the Fourier basis. Graph Attention Network (GAT)~\cite{velivckovic2017graph} aggregates neighbor nodes through multi-head self-attention mechanism, realizing the adaptive matching of the weights of different neighbors and enhancing the ability of Graph Convolutional Network (GCN). DisenGCN~\cite{ma2019disentangled} proposes a disentangled graph convolutional network with neighborhood routing mechanism to learn disentangled node representations from its neighbors. CGAT~\cite{lin2020graph} enhances the GAT framework by introducing a channel-aware attention mechanism. It disentangles topic representations structurally and semantically over user-user interaction graphs. 
	
	Graph neural network is also widely used in recommender systems. LightGCN~\cite{he2020lightgcn} learns the user and item embeddings by linearly propagating them on the user-item interaction graph. FGNN~\cite{qiu2019rethinking} investigates the inherent order of item transition patterns in session recommendation using a modified weighted GAT model. These existing studies have proved the effectiveness of graph neural network in obtaining item-link transition patterns, so we apply it to our work to model the user intention transition through sequences.
	
	\section{Preliminary}\label{preliminary}
	
	We will present the preliminary statements of this paper before the details of our model. We first describe the notations and the sequential recommendation problem of our paper. Then we put forward the channel-aware mechanism used in the representation space. Finally, we introduce the variational auto-encoder and its contribution to learning disentanglement.
	
	\subsection{Problem Formulation}
	
	Given $M$ users and $N$ items, we denote a user set as $U=\{u_1,u_2,...,u_M\}$ and an item set as $V=\{v_1,v_2,...,v_N\}$. For each user, $S^u=\{h_1,h_2,...,h_{|S^u|}\}$ represents the sequential behaviors interacted by user $u$. Given a historical sequence at time $t$, a sequential recommender model aims to predict the next item at time $t+1$.
	
	In this paper, we propose a global-level graph to capture item-link transition information. We define the global graph as $G=<V, E>$, where $V$ is a set of all items in the training data and $E$ is a set of edges. Each edge $<v_i,v_j> \in E$ means a user interacts with $v_j$ after $v_i$ in a sequence. $\mathcal{N}_{v_i}$ denotes the neighborhood of item $v_i$, i.e., the items adjacent to $v_i$ in the sequences. We use an undirected graph in this paper, because for the item-link pairs, the similarities and influencing factors between them are order-independent. The temporal order of sequences will be considered at the local level. Table~\ref{tab:notation} lists detailed explanations of the notations used in this paper.
	
	\begin{table}[t]
		\caption{Details of notations}
		\newcommand{\tabincell}[2]{\begin{tabular}{@{}#1@{}}#2\end{tabular}}
		\renewcommand\arraystretch{1.3}
		\centering
		\begin{tabular}{c|l}
			\hline
			\textbf{Notations}  &\textbf{Descriptions}\\ \hline
			$U,V$ &set of users and items \\
			$M,N$ &number of users and items\\
			$S^u$ &historical behaviors of user $u$ \\
			$K$ &number of decomposed channels \\
			$T,L$ &length of sequences and sliding windows \\
			$G=<V, E>$ &global item-link graph \\
			$\mathcal{N}_{v_i}$ &set of item $v_i$’s neighbors \\
			$d_{in},d_{channel}$ &embedding dimensions of inputs and channels \\
			$h_i,p_i$ &initial embedding of items and positions \\
			$\alpha_{i,j}^{(k)}$ &\tabincell{l}{probability of the correlation between item $v_i$ \\and $v_j$ regarding channel $k$} \\
			$z_i^{(k)}$ &representation of item $v_i$ regarding channel $k$ \\
			$z^g,z^l$ &representation of global and local level layers \\
			$z^s,z^v$ &output of self-attention and VAE layer \\
			$\mu^v,\sigma^v$ &\tabincell{l}{mean and variance of Gaussian distribution in \\VAE layer} \\
			\hline
		\end{tabular}
		\label{tab:notation}
	\end{table}
	
	\subsection{Channel-aware Mechanism}
	
	Given a historical sequence $S^u=\{h_1,h_2,...,h_{|S^u|}\}$, let $z_i \in \operatorname{R}^d$ be the latent intention of user $u$ while interacting with item $h_i$. Assuming that there are $K$ factors related to user intentions, we divide the latent representation $z_i$ into $K$ channels, i.e., $z_i=[z_i^{(1)},z_i^{(2)},...,z_i^{(K)}]$. The $k^{th}$ channel corresponds to the $k^{th}$ factor independently. For each pair of adjacent items, the correlation between $z_i^{(k)}$ and $z_j^{(k)}$ indicates the similarity between item $v_i$ and item $v_j$ regarding factor $k$, and also reveals why the two items are connected and how they influence each other.
	
	\subsection{VAE for Disentangled Learning}\label{preVAE}
	
	The Variational Auto-encoder (VAE) is a generative model which models variables as random distributions based on the Bayesian Theorem. Assume a $d$-dimensional variable $z$ being the sampled latent representation from sequence $S^u=\{h_1,h_2,...,h_{|S^u|}\}$, we aim to maximize the probability of the next item, that is, to maximize the probability of the whole sequence $S^u$:
	\begin{equation}
	p(S^u)=\prod_{h_t \in S^u} p(h_t|h_1,h_2,...,h_{t-1}).
	\end{equation}
	
	Since the probability $p(S^u)$ is not iterable, the variational inference method takes advantages of Bayesian Theorem $p(x,z)=p(x|z)p(z)$ and proposes a posterior distribution $q(z|x)$ to approximate the true distribution $p(z|x)$. Migrating to sequential recommendation, the log likelihood of $p(S^u)$ can be derived as follows:
	\begin{equation}
	\begin {aligned}
	\log p(S^u) &= \int q(z|S^u) \log p(S^u)  \mathrm{d}z \\
	&= \int q(z|S^u) \log  \frac{p(S^u,z)}{p(z|S^u)}  \mathrm{d}z \\
	&= \int q(z|S^u) [\log  \frac{p(S^u,z)}{q(z|S^u)} - \log \frac{p(z|S^u)}{q(z|S^u)}] \mathrm{d}z \\
	&\geq \int q(z|S^u) \log  \frac{p(S^u,z)}{q(z|S^u)} \mathrm{d}z \\
	&= \int q(z|S^u) \frac{p(S^u|z)}{q(z|S^u)} \mathrm{d}z + \int q(z|S^u) \frac{p(z)}{q(z|S^u)} \mathrm{d}z \\
	&= E_{z\sim q}[\log p(S^u|z)] - KL[q(z|S^u)||p(z)]\\
	\end {aligned}
	\label{ELBO}
	\end{equation}
	
	Algorithm~\ref{ELBO} is the training objective of variational auto-encoder, it is called Evidence Lower BOund (ELBO). By maximizing ELBO, the model can get an approximate posterior distribution $q(z|S^u)$ for the encoder to generate the latent representation $z$.
	
	In practice, the generative model suggests that the variables follow Gaussian distribution and applies a 'Reparameterization Trick' to calculate the gradient. Then the variables can be written as a polynomial generated from the mean $\mu$ and the variance $\sigma$ of Gaussian distribution:
	\begin{equation}
	z=\mu+\sigma\cdot\epsilon.
	\end{equation}
	
	$\beta$-VAE is a common modification of VAE. It introduces an adjustable hyperparameter $\beta$ to the original objective of VAE:
	\begin{equation}
	\mathrm{ELBO}=E_{z \sim q}[\log p(S^u|z)]-\beta KL(q(z|S^u)||p(z)).
	\end{equation}
	
	Burgess et.al.~\cite{burgess2018understanding} discussed why $\beta$-VAE is able to learn an axis-aligned disentangled representation from the perspective of information bottleneck. $\beta$ acts as a constriction limiting the capacity of the bottleneck, and encourages $\beta$-VAE to improve data log-likelihood.
	
	Furthermore, the KL-divergence term can be composed to three parts, following the contribution of $\beta$-TCVAE~\cite{chen2018isolating}:
	\begin{equation}
	\begin {aligned}
	&E_{z \sim q}[KL[q(z|S^u)||p(z)] = \iint q(S^u)q(z|S^u) \log \frac{q(z|S^u)}{p(z)} \mathrm{d}z\\
	&= \iint q(S^u)q(z|S^u) \log \frac{q(z|S^u)}{q(z)} \mathrm{d}z + \int q(z) \log \frac{q(z)}{p(z)} \mathrm{d}z\\
	&= I(z,S^u)+KL(q(z)||\prod_{j}p(z_{j})) + \sum_{j}KL(q(z_{j})||p(z_{j}))
	\end {aligned}
	\end{equation}
	
	The three terms above are referred to as the Mutual Information (MI), the Total Correlation (TC), and the dimension-wise KL respectively. A heavier penalty on the TC term forces the model to learn a factorized representation, each dimension of which is independent. Therefore, if we put a strong penalty on the KL-divergence by adjusting $\beta$, VAE can find statistically independent factors from the observed data.

	\section{Methodology}\label{methodology}
	
\begin{figure*}[thbp]
		\includegraphics[width=\textwidth]{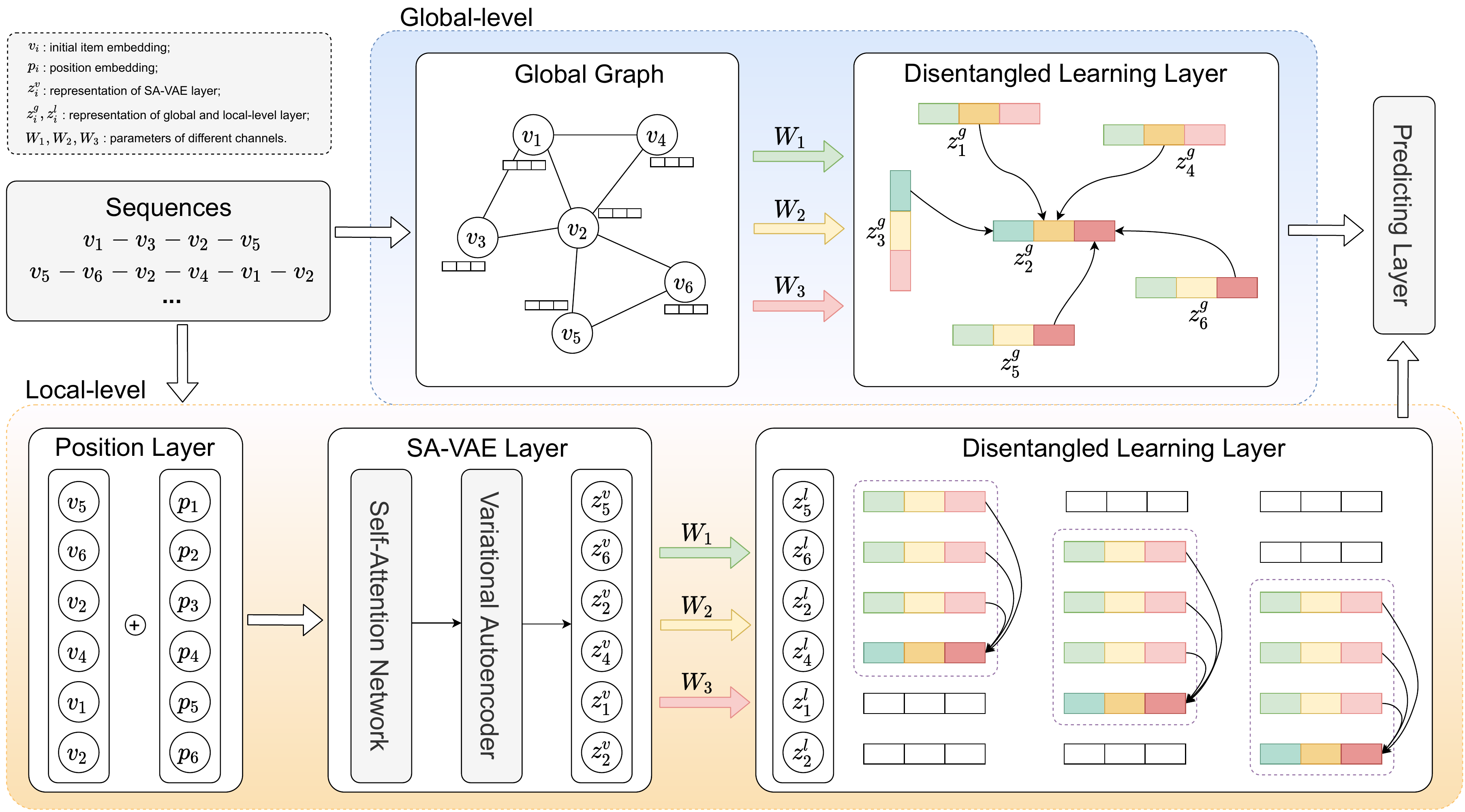}
		\caption{Overall architecture of the proposed model. The sequential representation is combined with the global representation $z^g$ learned from the Global-level Disentangled Representation Learning Layer and the local representation $z^l$ learned from the SA-VAE Layer and Local-level Disentangled Learning Layer.}
		\label{model}
	\end{figure*}

	In this section, we will present our proposed model. Figure~\ref{model} illustrates its overall architecture, which consists of three parts: Global-level Disentanglement Layer, Local-level Disentanglement Layer, and Prediction Layer. We claim that sequential disentangled representation learning model should have three main characteristics: 1) items that have similar features should be close in the corresponding embedding space; 2) the changing of factors between linked items should reveal the intention transition of user behaviors; 3) separated representations should be independent of each other. We will discuss how the model realizes these purposes in detail in the following sections.
	
	\subsection{Global-level Disentanglement Layer}
	
	We will first introduce the global-level disentangled representation learning layer based on the channel-aware mechanism, which is the key framework of our work. We build a global item-link graph $G=<V, E>$ based on training sequences, where all the item pairs appear adjacently in the sequences are connected with undirected edges. We aim to extract the independent factors motivating user intentions and find out the degree of mutual influence between the two items on these factors. Before introducing the mechanism, we first propose two hypotheses.
	
	\noindent \textbf{Hypothesis 1.} There are $K$ high-level concepts associated with user intentions, which means there are $K$ latent factors to be disentangled.
	
	Based on this hypothesis, given a global graph $G$, we divide the nodes (i.e., the items) into $K$ components in the latent space, and the edges are divided into $K$ channels correspondingly. The $k^{th}$ component is related to the $k^{th}$ factor of the user intention, and the $k^{th}$ channel indicates how factor $k$ attributes to the linkage of pairwise items.
	
	\noindent \textbf{Hypothesis 2.} Factor $k$ indicates the degree of similarity between item $v_i$ and $v_j$ in terms of factor $k$, that is, the representations of item $v_i$ and $v_j$ should be close in the $k^{th}$ latent subspace if $v_i$ and $v_j$ have similar characteristics regarding the $k^{th}$ factor.
	
	Intuitively, for a pair of linked items, their similarity is equivalent, which means on the same factor $k$, the degree of influence of item $v_i$ on $v_j$ is the same as that of $v_j$ on $v_i$. Therefore, we can use undirected graphs to model the information transition instead of directed graphs.
	
	\begin{figure}[t]
		\includegraphics[width=0.48\textwidth]{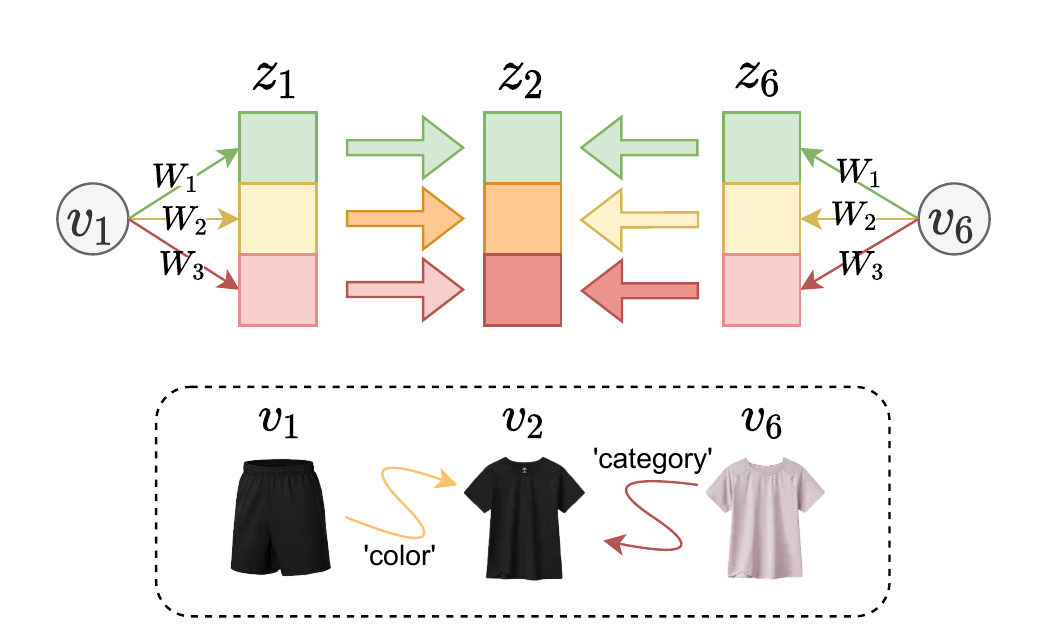}
		\caption{Illustration of channel-aware mechanism, where channel $W_2$ represents factor 'color' and channel $W_3$ represents factor 'category'.}
		\label{model_channel}
	\end{figure}
	
	Then we will introduce the channel-aware mechanism based on the above hypotheses. The illustration is shown in Figure~\ref{model_channel}. The item representations are divided into $K$ components by sending the initial embeddings into $K$ learning layers respectively. The edges are composed into $K$ channels and each channel transmits information of the corresponding item embedding. For a single node $v_i$ in the graph, we aim to aggregate information from its neighbourhood $\mathcal{N}_{v_i}$. We first compute the probability that factor $k$ influences item $v_i$ from its neighbors $v_j \in \mathcal{N}_{v_i}$:
	\begin{equation}
	\alpha_{i,j}^{(k)}=\dfrac{exp((\sigma(W_k^T h_i))^T\sigma(W_k^T h_j))}{\textstyle \sum_{k’=1}^{K}{exp((\sigma(W_{k’}^T h_i))^T\sigma(W_{k’}^T h_j))}},
	\label{equ:alpha}
	\end{equation}
	\noindent where $W_k \in \operatorname{R}^{d_{in} \times d_{channel}}$ is the parameter of the $k^{th}$ learning layer regarding factor $k$, and $h_i$ is the initial embedding of node $v_i$. $\sigma(\cdot)$ is a nonlinear activation function. $\alpha_{i,j}^{(k)}$ reveals why the item pair $v_i$ and $v_j$ is linked adjacently, and how item $v_j$ attributes to item $v_i$ over factor $k$. The larger $\alpha_{i,j}^{(k)}$ is, the higher item $v_i$ and $v_j$ are similar on factor $k$, the greater the transition information from $v_j$ to $v_i$ is, and the larger the width of the edge is in the graph. Moreover, $\alpha_{i,j}^{(k)}$ satisfies $\sum_{k’=1}^{K} \alpha_{i,j}^{(k’)} =1$ to ensure that the total width of each factor is the same.
	
	Then we can accumulate information according to the probabilities of channels from the neighbors of item $v_i$ and update the item representation:
	\begin{equation}
	z_i^{g(k)}=\sigma(W_k^T h_i)+\sum_{j \in \mathcal{N}_{v_i}}{\alpha_{i,j}^{(k)} (\sigma(W_k^T h_j))}.
	\label{equ:z}
	\end{equation}
	\noindent In order to ensure the numerical stability, we use $l_2$-normalization as:
	\begin{equation}
	z_i^{g(k)}=z_i^{g(k)}/\lVert z_i^{g(k)}\lVert _2.
	\end{equation} 
	
	By projecting item representations into different channels, we can aggregate item information from the perspective of different concepts. The global-level item representation $z_i^g$ can then be denoted as the combination of $K$ channels: 
	\begin{equation}
	z_i^g=[z_i^{g(1)}, z_i^{g(2)},..., z_i^{g(K)}].
	\end{equation}
	
	The design of channel-aware mechanism based on neighborhood fulfills our first claim of disentanglement learning. Similar characteristics are passed through corresponding channels to model item features and different neighbors influence the target item differently. Taking Figure~\ref{model_channel} as an example, item $v_1$ and $v_2$ are linked because they are black, then information will be passed through the channel which corresponds to factor 'color'. The representation of item $v_1$ and $v_2$ should be close in the component related to 'color', but far in the component related to 'category' since item $v_1$ is a pair of trousers and item $v_2$ is a shirt. Similarly, the representation of item $v_2$ and $v_6$ should be close in the component of 'category', but far in the component of 'color'.

	\subsection{Local-level Disentanglement Layer}
	
	Considering that items appearing in one sequence are rarely repeated, we model local-level user intentions based on sequential models instead of graphs. Given a sequence of user’s historical behavior $S^u=\{h_1,h_2,...,h_{|S^u|}\}$, we transform it into $S^u=\{h_1,h_2,...,h_T\}$ as training data. For users whose sequential length is greater than $T$, we select the nearest $T$ interacted items, and for those whose sequential length is less than $T$, we add zero vectors repeatedly to the left side of sequences. In order to distinguish the item representations at different positions in the sequence, we add a learnable position embedding $p \in \operatorname{R}^{T \times d_{in}}$ into the initial item embedding, and take $H$ as the input of the learning layer:
	\begin{equation}
	H=
	\begin{bmatrix}
	h_1 + p_1\\
	h_2 + p_2\\
	\cdots\\
	h_T + p_T
	\end{bmatrix}.
	\end{equation}
	
	\subsubsection{SA-VAE Layer}
	
	We first apply self-attention network ~\cite{kang2018self} into our local learning model taking advantage of its ability to capture both long and short-range dependencies of items in sequence. The scaled dot-product attention is defined following ~\cite{vaswani2017attention}:
	\begin{equation}
	D=Attention(Q,K,V)=softmax \left ( \frac{QK^T}{\sqrt{d_{in}}} \right )V,
	\end{equation}
	\noindent where $d_{in}$ is the dimension of input embedding, the scale factor $\sqrt{d_{in}}$ is to avoid the inner product values being overly large. $Q$, $K$, and $V$ denote the queries, keys and values respectively. The three parameters are generated by input $H$:
	\begin{equation}
	Q=HW^Q, K=HW^K, V=HW^V,
	\end{equation}
	\noindent where $W^Q, W^K, W^V \in \operatorname{R}^{d_{in}}$ are the projection matrices of attention layers.
	
	By utilizing residual connection and layer normalization, we can propagate low-level features to the high-level ones and get the final output of the self-attention layer:
	\begin{equation}
	h^s=LayerNorm(D + H).
	\end{equation}
	
	We then take $h^s$ as the input of the variational auto-encoder framework. Let $z^v$ be the latent variable sampled from the sequence $S^u$, which obeys a Gaussian distribution. Following SVAE~\cite{sachdeva2019sequential}, we inference the posterior distribution $q(z|S^u)$ as a multinomial layer. The mean and variance vectors are computed based on the self-attention vectors as follows:
	\begin{equation}
	\mu^v = l_1(h^s), \sigma^v = l_2(h^s),
	\end{equation}
	\noindent where $l(\cdot)$ represents linear transformations. By using the 'Reparameterization Trick' mentioned in the Preliminary section, the output of our SA-VAE layer is written as:
	\begin{equation}
	z^v=\mu^v+\sigma^v\cdot\epsilon,
	\end{equation}
	\noindent where $\epsilon \sim \mathcal{N}(0,I)$. By sampling a random variable $\epsilon$ with standard Gaussian distribution, the latent representation of sequence is reparameterized, and we can handle the uncertainty of user behaviors.
	
	\subsubsection{Disentangled Learning Layer}
	
	After the self-attention variational auto-encoder model, we get item representations with normal distribution of the overall sequence. Then, we will apply the channel-aware aggregation mechanism for local-level disentanglement learning.
	
	In the global-level learning layer, we assume that adjacent items have similar characteristics, so we apply the channel-aware mechanism to aggregate feature information for representation updating. When it comes to local level, we focus on the transition of user preference by modeling the variation of item factors. In order to obtain the transition features, we use the sliding window strategy based on graph neural network.
	
	Sliding window strategy is a popular dividing algorithm. By applying sliding window strategy to some search tasks, it can convert the nested loop problem into a single loop problem, reducing time complexity. Specifically, the algorithm sets a fixed window, which moves from time 1 to time $T$ in the sequence axis, and executes the channel-aware algorithm in each step among the window. Taking Figure~\ref{model} as an example, the user sequence is arranged by time order as $v_5-v_6-v_2-v_4-v_1-v_2$. Suppose the window length is 4, the window first covers the 4 items of the earliest interactions. We add edges between $v_5-v_4$, $v_6-v_4$ and $v_2-v_4$ respectively, and apply the channel-aware mechanism to calculate the similarities and degrees of influence between the three items and $v_4$ over the $K$ channels. Then we accumulate the feature information to $v_4$, achieving a step of information transmission. Next, the window slides to $v_6-v_2-v_4-v_1$, and we repeat the above steps in this window. Finally, the item feature information will be transformed to the last item through $K$ channels.
	
	We set the sliding window length as $L$. That means for each target item $v_i$, information will be aggregated from its former $L$ items. The probability between item $v_i$ and $v_j$ is calculated by Algorithm~\ref{equ:alpha} based on the channel-aware mechanism and channel information is aggregated as follows:
	\begin{equation}
	\begin{aligned}
	\alpha_{i,j}^{(k)}=\dfrac{exp((\sigma(W_k^T z^v_i))^T\sigma(W_k^T z^v_j))}{\textstyle \sum_{k’=1}^{K}{exp((\sigma(W_{k’}^T z^v_i))^T\sigma(W_{k’}^T z^v_j))}},\\\\
	z_i^{l(k)}=\sigma(W_k^T z^v_i)+\sum_{L-i \leq j < i}{\alpha_{i,j}^{(k)} (\sigma(W_k^T z^v_j))},
	\end{aligned}
	\end{equation}
	\noindent where $W^k$ represents the learning parameter of channel $k$, which is shared with the global-level layer. Also, we use $l_2$-normalization for $z_i^l$. Having obtained the information aggregation through sliding window from previous to back, we can form the user intention at time $t$ with the $t^{th}$ item embedding in the sequence. Then the local sequential representation is the combination of $K$ factors: $z_t^l=[z_t^{l(1)}, z_t^{l(2)},..., z_t^{l(K)}]$.
	
	In summary, we learn disentangled representation of the current sequence from both channel and statistical perspectives. Variational auto-encoder helps the model learn independent latent representation over the whole sequence statistically, which would be discussed in the next part. The channel-aware sliding window strategy is able to distinguish the various factors of users. Different factors pass through channels that are related to different user intentions. The influenced factor is changed through sequences with the transition of user intentions, realizing characteristic (2) of our claim.
	
	\subsection{Predicting Layer}
	
	Based on the obtained representations $z^{g}, z^{l}$ learned from global and local level layers, the final sequential representation is written as: 
	\begin{equation}
	z = W^g z^g+ W^l z^l,
	\end{equation}
	\noindent where $W^g, W^l$ are combination parameters of predicting layer. 
	
	We can estimate the final recommendation probability of candidate items based on the current sequential embedding and the initial item embedding. Let $\hat y_i$ denote the prediction probability of item $v_i$ appearing as the next interaction in the current sequence:
	\begin{equation}
	\hat y_i = softmax(z^T h_i).
	\end{equation}
	
	Since we use VAE framework in our model, the training objective is defined following the evidence lower bound:
	\begin{equation}
	\mathrm{ELBO}=E_{q(z|S^u)}[\log p(S^u|z)]-\beta KL(q(z|S^u)||p(z)).
	\end{equation}
	
	The first term $E_{q(z|S^u)}[\log p(S^u|z)]$ is regarded as the reconstruction error, measuring the accuracy between the prediction $\hat y_i$ and ground truth $y_i$. Here we compute the reconstruction loss using cross entropy:
	\begin{equation}
	E_{z \sim q}[\log p(S^u|z)]=-\sum_{i=1}^{N}{y_i \log (\hat y_i)+(1-y_i)\log (1-\hat y_i)}.
	\end{equation}
	
	The second term, $KL(q(z|S^u)||p(z))$, is used to measure the distance between posterior distribution and prior distribution. Practically, it is computed with the intermediate variables of VAE layer~\cite{kingma2013auto}:
	\begin{equation}
	KL(q(z|S^u)||p(z)) = \frac{1}{2} \sum_{i=1}^{N}(\mu_i^2+\sigma_i^2-\log  \sigma_i^2-1).
	\end{equation}
	
	We then discuss how our model can learn independent disentangled representation. According to Section~\ref{preVAE}, the KL-divergence $KL(q(z|S^u)||p(z))$ can be separated into three parts: the index-code mutual information, the total correlation, and the dimension-wise KL. The total correlation $KL(q(z) \lVert \prod_{j}{q(z_j)})$ is a measure of redundancy, acting as the degree of interdependence between variables in the latent variable space. Therefore, applying the $\beta$-VAE framework into our model contributes to learning statistically independent factors of the data distribution, and realizing our last claim of learning disentangled representation. 
	
	\begin{algorithm}[t]
		\caption{Channel-aware Algorithm} 
		\hspace*{0.02in} {\bf Input:} 
		initial item embeddings $h_i$, channel number $K$\\ 
		\hspace*{0.02in} {\bf Output:} 
		disentangled item embeddings $z_i$\\
		\hspace*{0.02in} {\bf Parameters:}
		$W_k, k=1,2,...,K$
		\begin{algorithmic}[1]
			\For{each item $v_i$} 
			\State $z_i^{(k)}=\sigma(W_k^T h_i)$
			\EndFor
			\For{each item $v_i$}
			\For{each item $v_j \in \mathcal{N}_{v_i}$}
			\State $\alpha_{i,j}^{(k)}=\dfrac{exp(z_i^{(k)^T} z_j^{(k)})}{\textstyle \sum_{k’=1}^{K}{exp(z_i^{(k')^T} z_j^{(k')})}}$
			\State $z_i^{(k)}=z_i^{(k)}+\alpha_{i,j}^{(k)}*z_j^{(k)}$
			\EndFor
			\EndFor
			\For{each item $v_i$} 
			\State $z_i^{(k)}=z_i^{(k)}/\lVert z_i^{(k)}\lVert _2$
			\EndFor
			\State $z_i=[z_i^{(1)}, z_i^{(2)},..., z_i^{(K)}]$
			\State \Return $z_i$
		\end{algorithmic}
		\label{alg 1}
	\end{algorithm}
	
	\subsection{Complexity Analysis}
	
	\subsubsection{Time Complexity}
	
	The time consumption of our model mainly consists of three parts. The first part is the global-graph building. In order to construct the global, we need to traverse every edge, which costs $O(E)$. The second part is the channel-aware mechanism, the algorithm of the mechanism is shown in Algorithm~\ref{alg 1}. For each channel, the cost of updating item embedding is $O(N^2d_{in}d_{channel})$. The third part is sliding window strategy. The window slides from the start of sequences to the end, which costs time of sequence length $T$, and the local-level channel-aware mechanism costs $O(LTd_{in}d_{channel})$. Therefore, the total space complexity of our model is $O(E+K(N^2+LT)d_{in}d_{channel})$. 
	
	\subsubsection{Space Complexity}
	
	The space consumption of our model is mainly in the undirected graph and channel-aware mechanism. In order to store the neighborhood of each node, an adjacency matrix of $O(N^2)$. For the channel-aware mechanism, there are $K$ parameters of dimension $d_{in}d_{channel}$ for all channels, and the final item embedding generated from $K$ channels is of dimension $NTd_{channel}$. Therefore, the total space complexity of our model is $O(N^2+Kd_{in}d_{channel}+KNTd_{channel})$.

	\section{Experiment}\label{experiment}
	
	In this section, we will present our experimental setup and results. Firstly, we introduce the datasets and the evaluation metrics used in our experiments, then we will introduce the eight baseline methods which are related to VAE models or disentangled learning models. Next, we compare the experimental results of these baseline methods with our method under the same experimental setting to verify the effectiveness of our proposed model. Moreover, we evaluate the influence of each part in our model and the influence of the key parameters. Finally, we perform visualization experiments on the sequential embeddings generated in the experiment, which proves that our disentanglement model is able to distinguish intention factors in the latent space. In specific, our experiments aim to answer the following questions:
	
	\begin{itemize}
		\item RQ1: Does our proposed model outperform the state-of-art works over various kinds of datasets?
		
		\item RQ2: What is the influence of each component, i.e., the global and local layers in our model?
		
		\item RQ3: How does our model realize the disentangled representation learning in the latent space?
		
		\item RQ4: What is the influence of the hyperparameter setting on different datasets?
	\end{itemize}
	
	\subsection{Datasets} 
	
	\begin{table}[t]
		\caption{Details for three datasets}
		\newcommand{\tabincell}[2]{\begin{tabular}{@{}#1@{}}#2\end{tabular}}
		\renewcommand\arraystretch{1.4}
		\renewcommand\tabcolsep{8pt}
		\centering
		\begin{tabular}{c|c|c|c|c}
			\hline
			\textbf{Dataset} &  \textbf{Users} & \textbf{Items} & \textbf{Interacts} & \textbf{Density}\\ \hline
			ML-1M        & 6040   & 3416   & 999611   & 95.15$\%$ \\
			Beauty      & 52204  & 57289  & 394908   & 99.98$\%$ \\
			Games       & 31013  & 23715  & 287107   & 99.96$\%$ \\
			\hline
		\end{tabular}
		\label{data} 
	\end{table}
	
	We adopt three real-world datasets to evaluate the effectiveness of our method. MovieLens is a time-series dataset containing rating data for multiple movies by users. We use the version MovieLens-1M that includes 1 million user ratings. Amazon is an e-commerce dataset which contains users’ purchasing behaviors on rich products. We choose two categories, 'Beauty' and 'Video Games', and use the 5-core version for our experiment. 
	
	We use timestamps to arrange the sequence order, that is, the items that are interacted by the same users are arranged in sequence according to their interacting time. Following the previous work~\cite{kang2018self}, we split data into three parts: the last interacted item for testing, the second-to-last interacted item for validation and the rest items for training. We regard the training sequence of length $n$ as $n-1$ sub-sequences, and the last element of each sub-sequence is regarded as the training ground truth. While in validation and testing tasks, we choose the last item of sequence as ground truth with 100 randomly sampled negative items. The detailed statistics are shown in Table~\ref{data}. The average sequence length of each dataset is 163.5, 5.63 and 7.26 respectively.
	
	\subsection{Metrics} 
	
	We adopt two ranking based metrics to evaluate the recommendation performance: Normalized Discounted Cumulative Gain (NDCG) and Recall. The larger the values of metrics are, the better the performance is. We refer to the two metrics as N@K and R@K for short.
	
	\begin{itemize}
		\item NDCG is a rating metric which takes into account the position of correctly recommended items. It is defined as follows:
		\begin{equation}
		NDCG@K = \frac{DCG@K}{IDCG@K},
		\end{equation}
		where DCG is the Discounted Cumulative Gain. We hope that the most relevant items are at the top of the list, so before adding scores, we divide each item by an increasing number. IDCG is the ideal DCG, which sorts the results to the best state and calculates DCG of the query under this arrangement. They are defined as:
		\begin{equation}
		\begin {aligned}
		DCG@K &= \sum_{i=1}^{K}\frac{2^{r_i}-1}{\log_2(i+1)},\\
		IDCG@K &= \sum_{i=1}^{N}\frac{1}{\log_2(i+1)},
		\end {aligned}
		\end{equation}
		\noindent where $r_i$ represents the relevance of the $i^{th}$ item, which is either 1 or 0, and $N$ is the set of relevant items.
		
		\item Recall describes the percentage of rated items that are actually preferred by users included in the recommendation list. It defines a recommendation list of top $K$ predicted items for a user as $R_K$, and uses $T$ to represent the corresponding test set. The percentage of rated items is then computed as:
		\begin{equation}
		Recall@K = \frac{\lvert T\cap R_K \rvert}{\lvert T \rvert}.
		\end{equation}
	\end{itemize}
	
	\begin{table*}[t]
		\caption{Results of recommendation performance}
		\newcommand{\tabincell}[2]{\begin{tabular}{@{}#1@{}}#2\end{tabular}}
		\renewcommand\arraystretch{1.4}
		\renewcommand\tabcolsep{6pt}
		\centering
		\begin{threeparttable}
			\begin{tabular}{c|c|ccccccccc}
				\hline
				\textbf{Dataset}        & \textbf{Metric} & \textbf{POP} & \textbf{BPR} & \textbf{FPMC} & \textbf{TransRec} & \textbf{Caser} & \textbf{SASRec} & \textbf{DSS}\tnote{1} & \textbf{VSAN}   & \textbf{EGD-GNN} \\ \hline
				\multirow{4}{*}{ML-1M}  & N@5             & 0.1428       & 0.1856       & 0.2726        & 0.2816            & 0.2175         & 0.3922          & 0.2119       & \underline{0.4224}    & \textbf{0.4571}  \\
				& R@5             & 0.2546       & 0.2972       & 0.4081        & 0.4166            & 0.3389         & 0.5450          & 0.3202       & \underline{0.5851}    & \textbf{0.6161}  \\
				& N@10            & 0.1863       & 0.2440        & 0.3284        & 0.3352            & 0.2709         & 0.4419          & 0.2562       & \underline{0.4593}    & \textbf{0.5012}  \\
				& R@10            & 0.4086       & 0.4738       & 0.5806        & 0.5826            & 0.5045         & 0.6982          & 0.4573       & \underline{0.7054}    & \textbf{0.7517}  \\ \hline
				\multirow{4}{*}{Beauty} & N@5             & 0.0483       & 0.0915       & 0.1429        & 0.1726            & 0.1928         & 0.2166          & 0.2139       & \underline{0.2285}    & \textbf{0.2380}  \\
				& R@5             & 0.0754       & 0.1299       & 0.2220        & 0.2467            & 0.2787         & 0.3013          & 0.3155       & \textbf{0.3332} & \underline{0.3158}     \\
				& N@10            & 0.0659       & 0.1448       & 0.1839        & 0.2049            & 0.2295         & 0.2495          & 0.2527       & \underline{0.2575}    & \textbf{0.2710}  \\
				& R@10            & 0.1303       & 0.2425       & 0.3492        & 0.3471            & 0.3923         & 0.4030          & 0.4351       & \textbf{0.4279} & \underline{0.4180}     \\ \hline
				\multirow{4}{*}{Games}  & N@5             & 0.1695       & 0.2195       & 0.2004        & 0.2428            & 0.2671         & \underline{0.3978}    & 0.2589       & 0.3956          & \textbf{0.4303}  \\
				& R@5             & 0.2845       & 0.3204       & 0.3250        & 0.3468            & 0.3821         & 0.5377          & 0.3714       & \underline{0.5491}    & \textbf{0.5599}  \\
				& N@10            & 0.2082       & 0.2606       & 0.2583        & 0.2845            & 0.3123         & \underline{0.4388}    & 0.3052       & 0.4288          & \textbf{0.4668}  \\
				& R@10            & 0.3605       & 0.4459       & 0.4462        & 0.4762            & 0.5217         & 0.6641          & 0.5152       & \underline{0.6672}    & \textbf{0.6723}  \\ \hline
			\end{tabular}
			\begin{tablenotes}
				\footnotesize
				\item[1] The code of DSS is not released by authors and we re-implement it according to the paper.
			\end{tablenotes}
		\end{threeparttable}
		\label{perf}
	\end{table*}
	
	\subsection{Baselines} 
	
	We compare our method with the following competitive baselines, with particular emphasis on VAE-based and disentangled learning methods. All the baselines put emphasis on sequential recommendation tasks. 
	
	\begin{itemize}
		\item 1) POP: a classical method that ranks items according to their popularity. 
		\item 2) BPR: Bayesian Personalized Ranking~\cite{rendle2012bpr}, a classical model based on Matrix Factorization. It designs a pair-wise optimization method to learn pairwise item rankings from implicit feedback. 
		\item 3) FPMC: Factorized Personalized Markov Chains~\cite{rendle2010factorizing}, a method combining Matrix Factorization and first-order Markov Chain together. It introduces a personalized transfer matrix based on Markov chain to capture time information and introduces matrix factorization to solve the sparse problem of the transition matrix.
		\item 4) TransRec: Translation based Recommendation~\cite{he2017translation}. It embeds items into a transition space and models each user as a transition vector to obtain the 'three-order' relationships, i.e., the interactions between a user, the previous visited items and the next item. 
		\item 5) Caser: Convolutional Sequence Embedding Recommendation~\cite{tang2018personalized}. The main idea is to form an 'image' with the most recent items of a sequence in time and latent spaces, and apply Convolutional Neural Network (CNN) to learn the high-order sequential patterns as the local feature of the image.
		\item 6) SASRec: Self-Attention based Sequential Recommendation~\cite{kang2018self}. By applying the self-attention mechanism into sequential problems, the model can not only capture the long term information like RNNs but also handle the short term patterns in terms of small number of behaviors like MCs. 
		\item 7) DSS: Disentangled Self-Supervision~\cite{ma2020disentangled}, the first model that focuses on disentangled representation learning on sequential recommendation. It designs a Disentangled Sequence Encoder to disentangle user intention in the latent space over sub-sequences and propose a seq2seq self-supervised strategy for training.
		\item 8) VSAN: Variational Self-attention Network~\cite{zhao2021variational}. It combines the self-Attention mechanism with variational inference for sequential recommendation to model the long-range and short dependencies of sequences. 		
	\end{itemize}
	
	\subsection{Experiment Setting} 	
	
	We conduct experiments with PyTorch. In the experiments, the dimension of item embedding of all the methods is set 100. The channel embedding dimension of our model is set 20. We set the batch size as 128 and the learning rate as 0.002. We limit the maximum sequence length to 200 for the MovieLens dataset and 50 for Amazon. The dropout rate of turning off neurons is set as 0.5 for both the global and local layers. Single-head self-attention network is used as the sequential encoder. We use random seeds for the generation of Gaussian distribution and report the average performance result under five times.
	
	\subsection{Performance Analysis}
	
	To evaluate the effectiveness of our proposed model, we perform next-item recommendation based on our model and the baselines under the same experimental setting. Specifically, we predict the item user may be interested in at time $t$ based on the former $t-1$ items and choose the $t^th$ item in each sequence as ground truth for metric. Table~\ref{perf} records the performance results. We will compare and analyze the results in detail in this section.
	
	Firstly, We can observe from the table that our method outperforms the baselines over all the datasets. There is no doubt that our model gets better predicting results over the classical baselines, POP, BPR, and FPMC, since we take complex sequential interaction information into account. In terms of models based on neural networks, we can see that SASRec performs better than the transformer models, indicating that the self-attention network captures more sequential semantics with both long and short term patterns.
	
	The VSAN method proposes a new self-attention network with variational auto-encoder and achieves second-best results in our experiments. It proves that capturing the long and short range dependencies together with the attention-based network and the statistical method does help the model get better prediction results. The effectiveness of random method, variational inference, is also confirmed in eliminating the random noise in user behaviors. Although the disentangled self-supervised method performs well in the Beauty dataset, it does not have good results in the other two datasets. However, its good performance on Beauty dataset is sufficient to prove the effectiveness of learning disentangled user intention over sequences. In DSS, one sequence behavior is encoded into one kind of user intention, ignoring the various different factors hidden behind item transitions. Compared with disentangling user intention over whole sequences, we focus on the intention transition of pairwise items, therefore, our model gets better predicting results than the previous work.
	
	Then we turn focus back to our proposed model, we get the best experimental results in most circumstances. In particular, its relative improvements over the strongest baselines w.r.t. NDCG@5 are 8.21$\%$, 4.16$\%$, and 8.17$\%$ for the three datasets respectively. Compared with VSAN, our model builds a global item-link graph and disentangles the influential factors into channels for representation updating. Compared with DSS, our model pays more attention to the item-item relationship. We form the user intention taking advantage of the transformer ability of self-attention instead of modeling the whole sequence with an encoder. According to these improvements, it is no doubt that our model can obtain user intention in an adaptive way and find more suitable items that users may be interested in.
	
	Secondly, we achieve the best improvements for all the metrics on MovieLens dataset. It indicates that by introducing the channel-aware mechanism, the model is able to capture more item-link relationship information that is hard to be captured by previous works. Moreover, by composing several high-level concepts, the movie items, which have few explicit features, is classified into some implicit categories, and the model can obtain the user intention from various high-level perspectives to predict users’ true preference.
	
\begin{figure}[t]
		\centering
		\subfigure[ML-1M]
		{
			\begin{minipage}[b]{.46\linewidth}
				\centering
				\includegraphics[width=\textwidth]{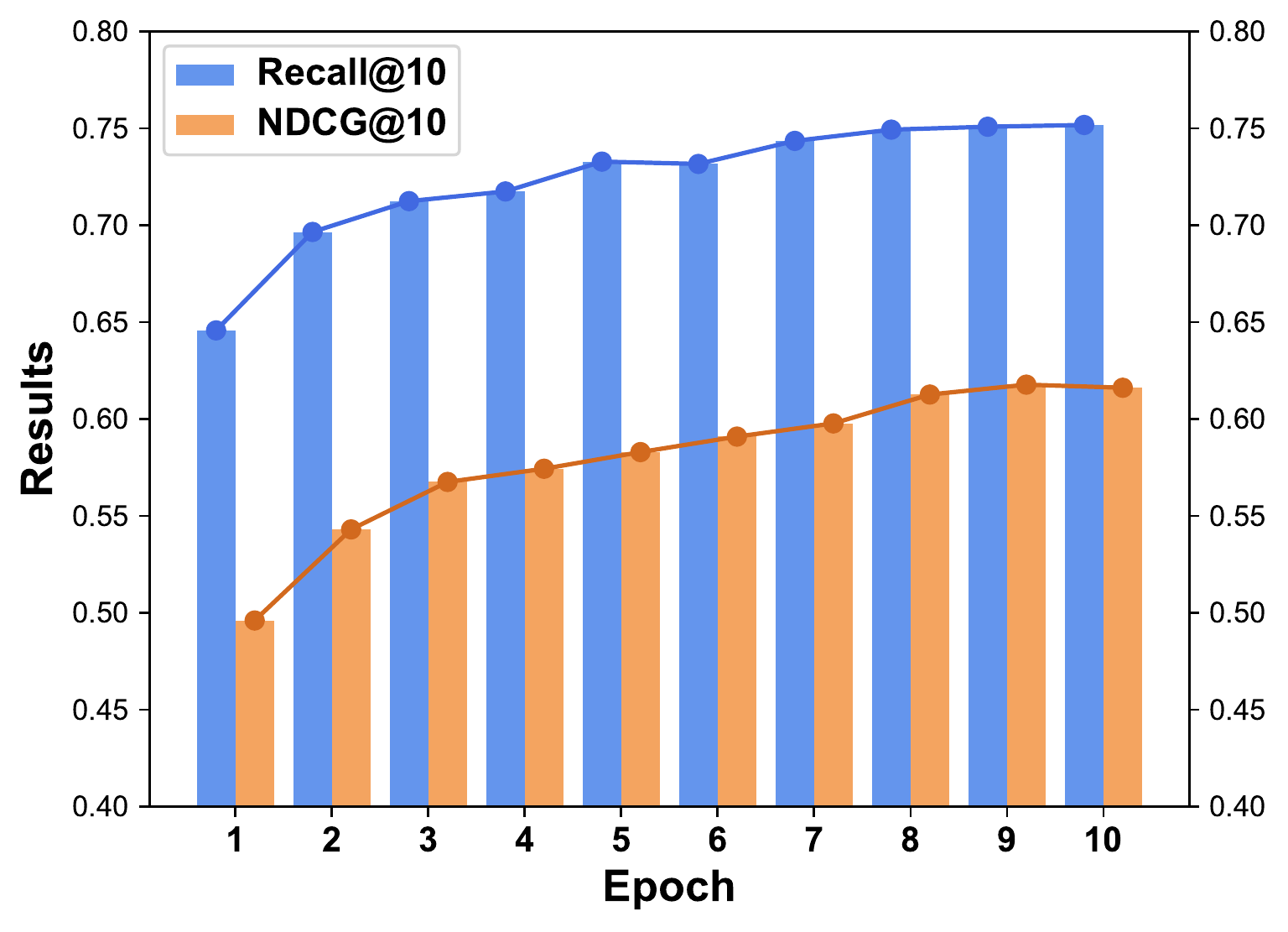}
			\end{minipage}
		}
		\subfigure[Beauty]
		{
			\begin{minipage}[b]{.46\linewidth}
				\centering
				\includegraphics[width=\textwidth]{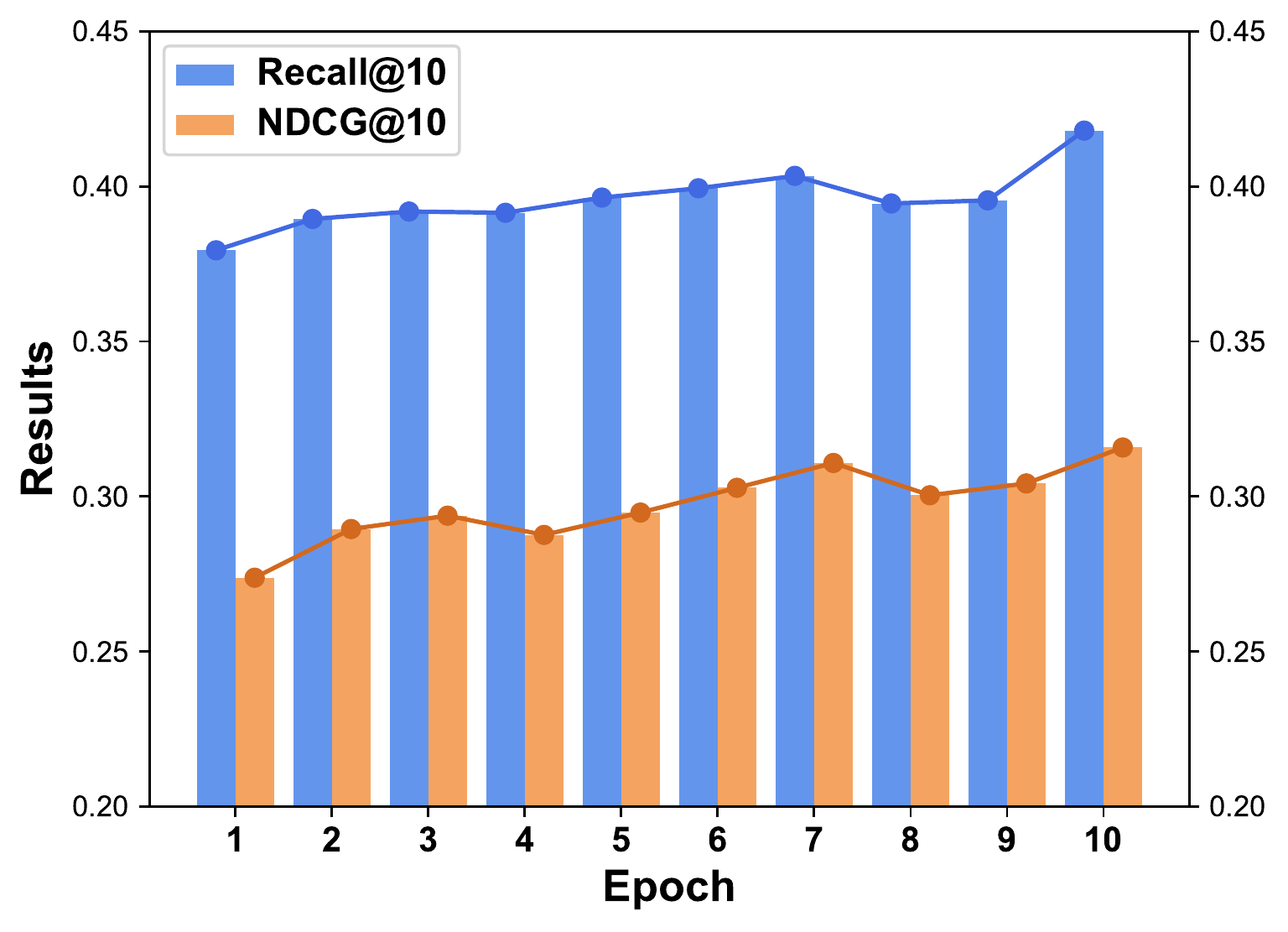}
			\end{minipage}
		}
		\caption{The trend of performance results in 10 epochs.}
		\label{speed}
	\end{figure}

	Thirdly, we find that our model reaches high values very early compared with the baselines, as shown in Figure~\ref{speed}. The trends of experimental results of 10 epochs indicate that our model can get good prediction results early in the first five epochs. Even though the time complexity of our model is larger than the state-of-art baselines, we can still get high prediction results within a short time. That means by distinguishing the latent factors hidden behind sequences, the model can learn item representations over various factors and find dynamic user intentions that are not shown explicitly.

	\begin{table}[t]
		\caption{Influence of each part of model}
		\newcommand{\tabincell}[2]{\begin{tabular}{@{}#1@{}}#2\end{tabular}}
		\renewcommand\arraystretch{1.5}
		\renewcommand\tabcolsep{4pt}
		\centering
		\begin{tabular}{c|c|ccccc}
			\hline
			\textbf{Dataset}        & \textbf{Metric} & \textbf{Global} & \textbf{Local} & \textbf{SA-VAE} & \textbf{SliWin} & \textbf{EGD-GNN} \\ \hline
			\multirow{2}{*}{ML-1M}  & N@10            & 0.4641          & 0.4772         & 0.2900          & 0.4628          & \textbf{0.5012}  \\
			& R@10            & 0.7142          & 0.7377         & 0.5311          & 0.7065          & \textbf{0.7517}  \\ \hline
			\multirow{2}{*}{Beauty} & N@10            & 0.2562          & 0.2437         & 0.2274          & 0.2420          & \textbf{0.2710}  \\
			& R@10            & 0.4043          & 0.4089         & 0.3990          & 0.4065          & \textbf{0.4180}  \\ \hline
			\multirow{2}{*}{Games}  & N@10            & 0.4445          & 0.3988         & 0.3370          & 0.3665          & \textbf{0.4668}  \\
			& R@10            & 0.6518          & 0.6242         & 0.5630          & 0.6050          & \textbf{0.6723}  \\ \hline
		\end{tabular}
		\label{abla}
	\end{table}
	
	\begin{table}[t]
		\caption{Influence of penalty on KL-divergence}
		\newcommand{\tabincell}[2]{\begin{tabular}{@{}#1@{}}#2\end{tabular}}
		\renewcommand\arraystretch{1.5}
		\renewcommand\tabcolsep{6pt}
		\centering
		\begin{tabular}{c|c|ccccc}
			\hline
			\textbf{Dataset}        & \textbf{Metric} & \textbf{$\boldsymbol{\beta}$=0} & \textbf{$\boldsymbol{\beta}$=0.1}    & \textbf{$\boldsymbol{\beta}$=0.5} & \textbf{$\boldsymbol{\beta}$=1}      & \textbf{$\boldsymbol{\beta}$=2} \\ \hline
			\multirow{2}{*}{ML-1M}  & N@10            & 0.4875     & \textbf{0.5012} & 0.4920       & 0.4881          & 0.4987     \\
			& R@10            & 0.7425     & \textbf{0.7517} & 0.7434       & 0.7478          & 0.7441     \\ \hline
			\multirow{2}{*}{Beauty} & N@10            & 0.2559     & 0.2587          & 0.2606       & \textbf{0.2710} & 0.2538     \\
			& R@10            & 0.3995     & 0.4078          & 0.4051       & \textbf{0.4180} & 0.4021     \\ \hline
			\multirow{2}{*}{Games}  & N@10            & 0.4472     & \textbf{0.4668} & 0.4625       & 0.4613          & 0.4616     \\
			& R@10            & 0.6671     & \textbf{0.6723} & 0.6699       & 0.6685          & 0.6659     \\ \hline
		\end{tabular}
		\label{abla-beta}
	\end{table}

	\subsection{Ablation Study}
	
	\subsubsection{Influence of each layer of model} 
	
	We first implement ablation studies to evaluate the effectiveness of each part of our model. Specifically, we perform four ablation experiments as follows:
	
	\begin{itemize}
		\item Global only: remove the local-level learning layer, i.e., the SA-VAE layer and sliding window layer, only perform with the global-level graph.
		\item Local only: remove the global link graph and corresponding learning layer, only perform with the local-level layer.
		\item SA-VAE only: remove the sliding window strategy part, only reserve the  self-attention and variational auto-encoder layers.
		\item SliWin only: remove the self-attention and variational auto-encoder layers, only reserve the sliding window mechanism for local-level learning.
	\end{itemize}
	
	Table~\ref{abla} lists the results of ablation studies, showing how each part influences the final performance of our model. It is clear that the global and local-level layers both contribute to the improvement of our model. For Amazon datasets, the global-level layer performs better than the local-level layer, indicating that the relationship between product items is close and worth exploring. 
	
	Besides, the sliding window strategy gets better prediction results compared with the SA-VAE layer. It proves that the channel-aware mechanism plays quite a crucial role in disentangling user intentions over different factors. Moreover, we can observe that the improvement of channel-aware mechanism is extremely large on the MovieLens dataset, since our model can capture much relevant information between items and explore high-level factors even on a small scale dataset.
	
	\subsubsection{Influence of penalty on KL-divergence}
	
	Then we implement ablation experiments on the variational auto-encoder framework. As introduced in the previous sections, the parameter $\beta$ acts as a penalty on KL-divergence term which contributes to forcing the model to find independent latent variables. Therefore, we will evaluate the role of $\beta$ in disentangled representation learning. We set $\beta$ from 0 to 2 to examine its effectiveness, and list the results in Table~\ref{abla-beta}. We can see that when $\beta$ is 0, the experimental results are obviously the worst, since the variational auto-encoder model degenerates to original auto-encoder. And when $\beta$ is too large, the results will also decrease, since the posterior distribution is close to the standard normal distribution. Therefore, we need to find a suitable value to strike a balance between reconstruction accuracy and disentangled learning.
	
	\begin{figure}[t]
		\centering
		\subfigure[Beauty]
		{
			\begin{minipage}[b]{.44\linewidth}
				\centering
				\includegraphics[width=\textwidth]{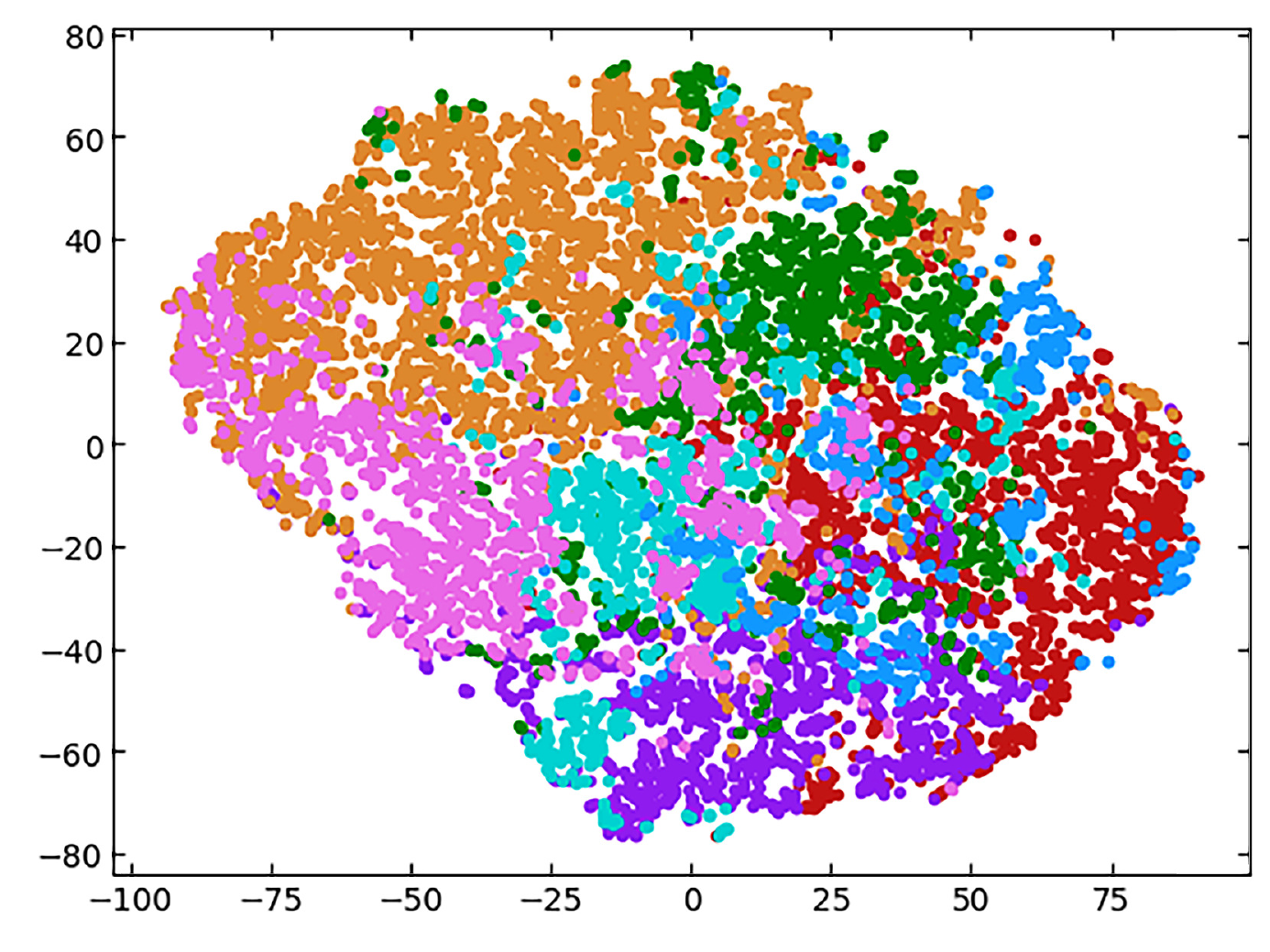}
			\end{minipage}
		}
		\subfigure[Games]
		{
			\begin{minipage}[b]{.45\linewidth}
				\centering
				\includegraphics[width=\textwidth]{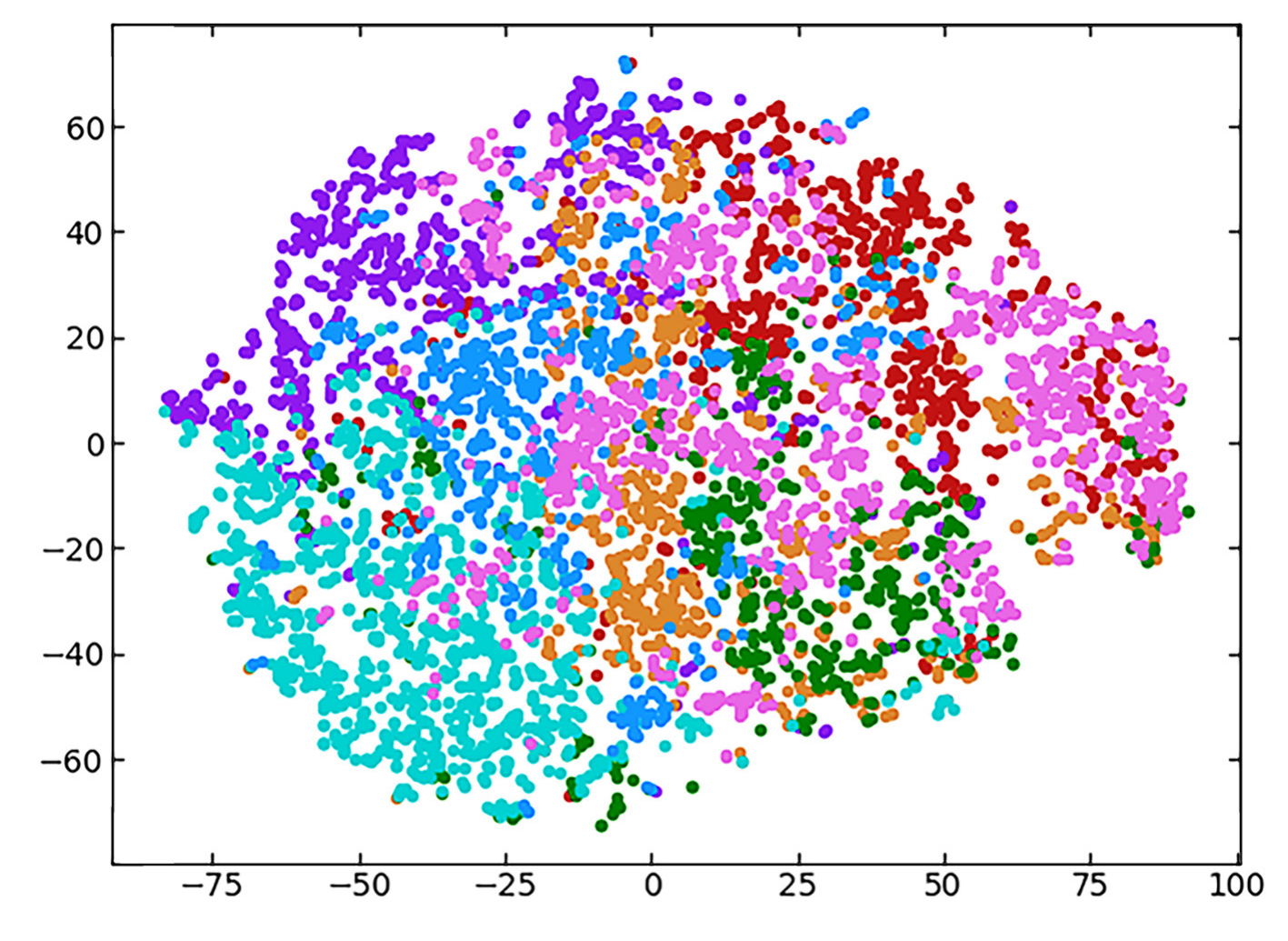}
			\end{minipage}
		}
		\caption{Visualization of the item embeddings on Amazon datasets}
		\label{visual}
	\end{figure}
	
	\subsection{Visualization of Node Representations}
	
	In order to analyze the performance of learning disentangled representation, we visualize the item representations using the t-SNE~\cite{van2008visualizing} algorithm on the two Amazon datasets. In detail, we learn the global-level item embeddings based on the channel-aware mechanism and project the embeddings into a 2-dimension space. We choose the channel with the largest embedding value, i.e., $max_k z_i^{g(k)}$, as the item category and color the nodes based on their categories in Figure~\ref{visual}.
	
	We can observe that on both datasets, the items with the same categories are close in the latent space, indicating they share similar features. Meanwhile, the factors which have close relationships are close in the latent space as well. Taking the Beauty dataset as an example, the items colored in pink are close to the items colored in orange and cyan. This indicates that the items which have these three features share similar characteristics. When a user interacts with an item colored pink, he/she is likely to choose an item colored in orange or cyan. The Games dataset also shows the same characteristics. This visual experiment again proves the effectiveness of learning disentangled representation from an intuitive perspective, and also shows its ability in enhancing the interpretability of model. In summary, learning disentangled representations based on item edges can not only observe the underlying features between items, but also help the model predict what users may like.

	\begin{figure*}[htbp]
		\centering
		\subfigure[ML-1M]
		{
			\begin{minipage}[b]{.315\textwidth}
				\centering
				\includegraphics[width=\textwidth]{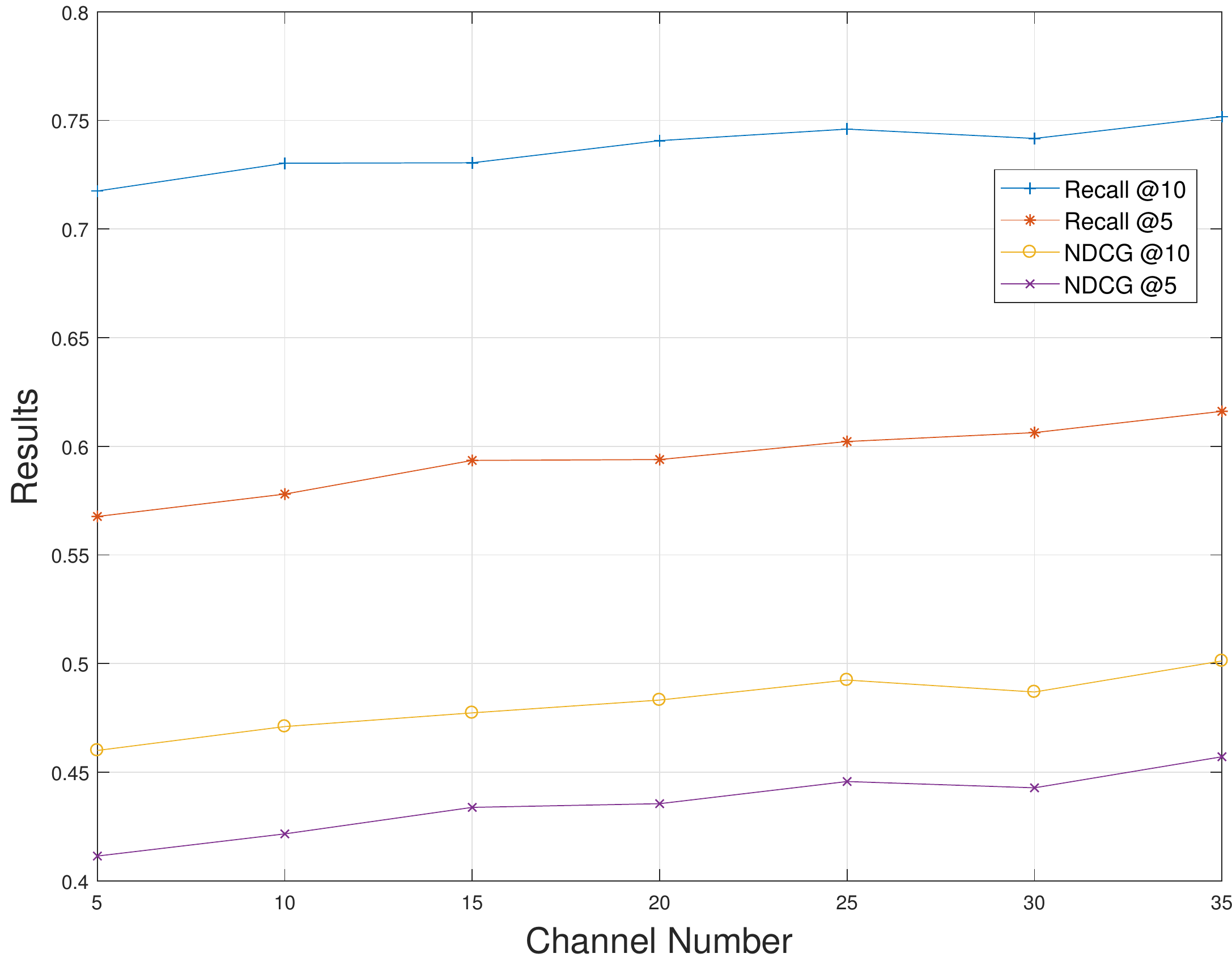}
			\end{minipage}
		}
		\subfigure[Beauty]
		{
			\begin{minipage}[b]{.315\textwidth}
				\centering
				\includegraphics[width=\textwidth]{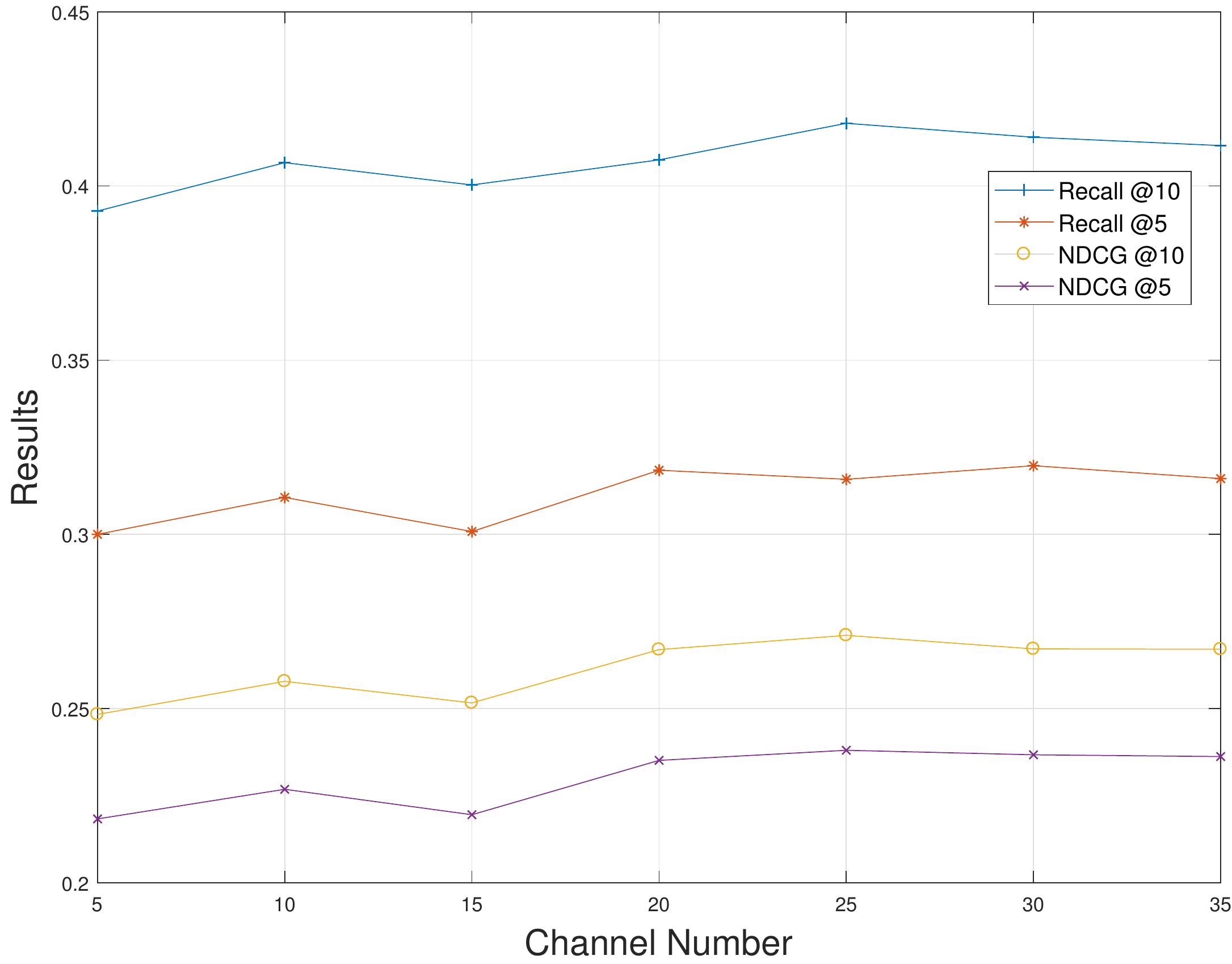}
			\end{minipage}
		}
		\subfigure[Games]
		{
			\begin{minipage}[b]{.315\textwidth}
				\centering
				\includegraphics[width=\textwidth]{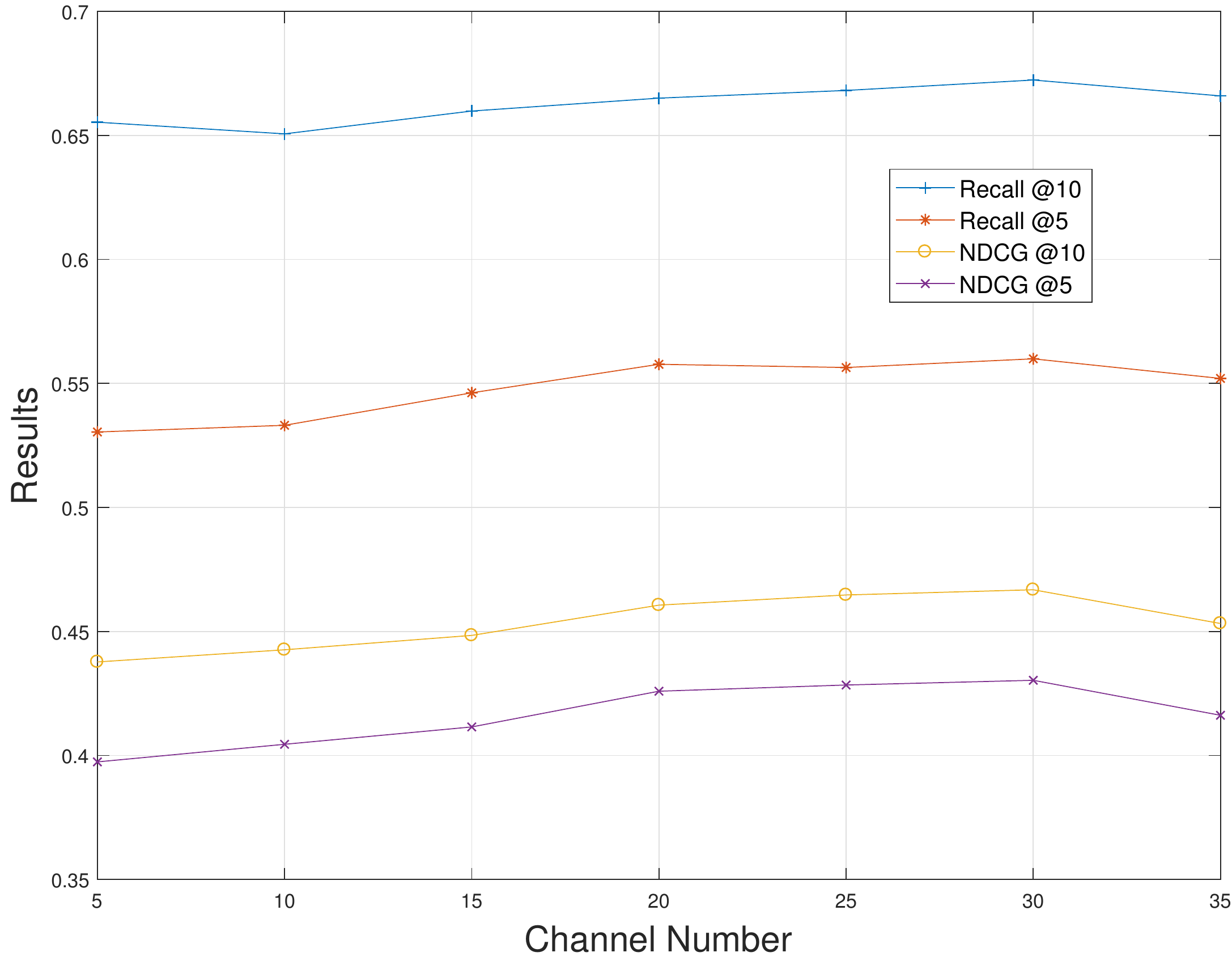}
			\end{minipage}
		}
		\caption{Effect of the number of channels}
		\label{channel}
	\end{figure*}
	
	\begin{figure*}[htbp]
		\centering
		\subfigure[ML-1M]
		{
			\begin{minipage}[b]{.315\textwidth}
				\centering
				\includegraphics[width=\textwidth]{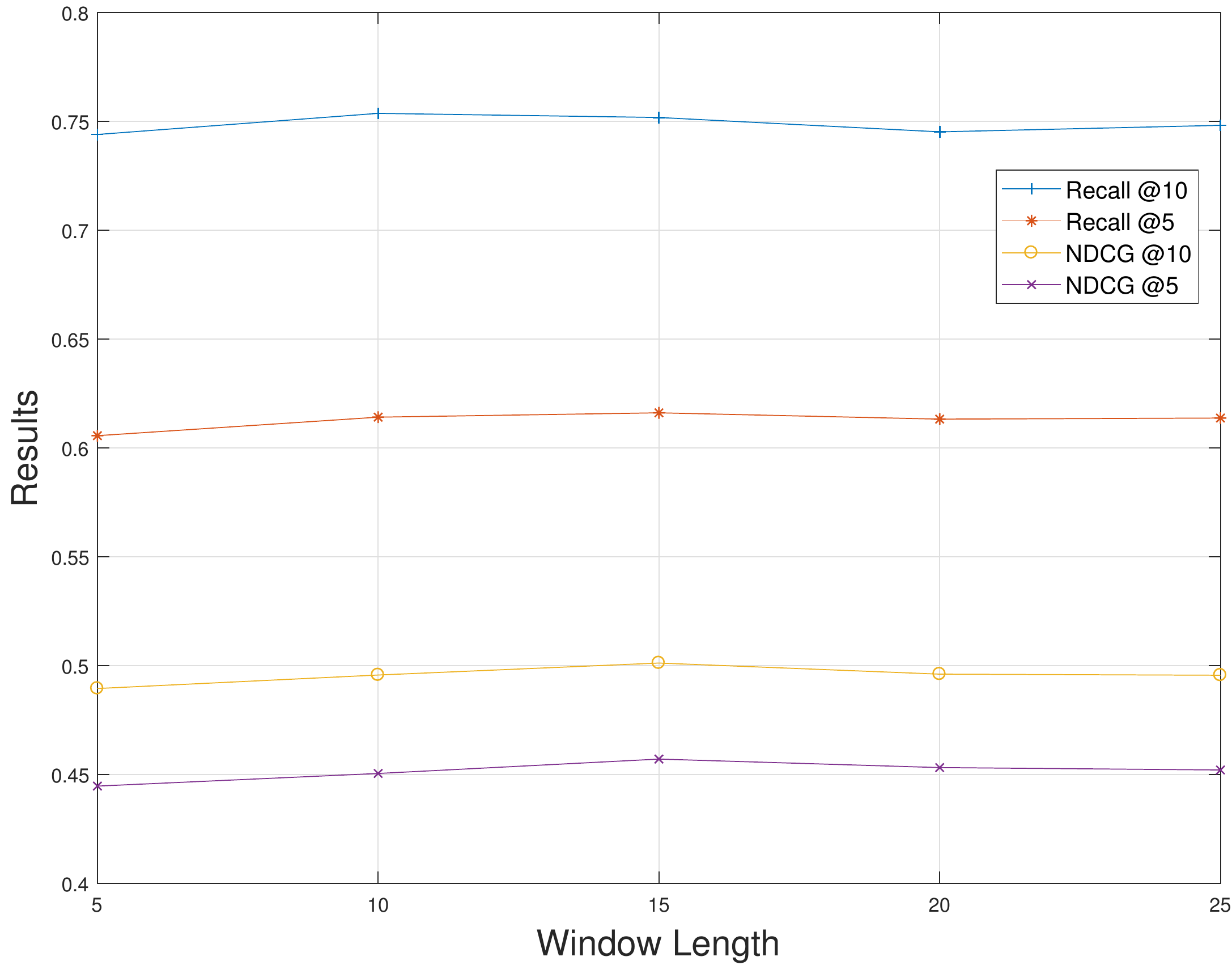}
			\end{minipage}
		}
		\subfigure[Beauty]
		{
			\begin{minipage}[b]{.315\textwidth}
				\centering
				\includegraphics[width=\textwidth]{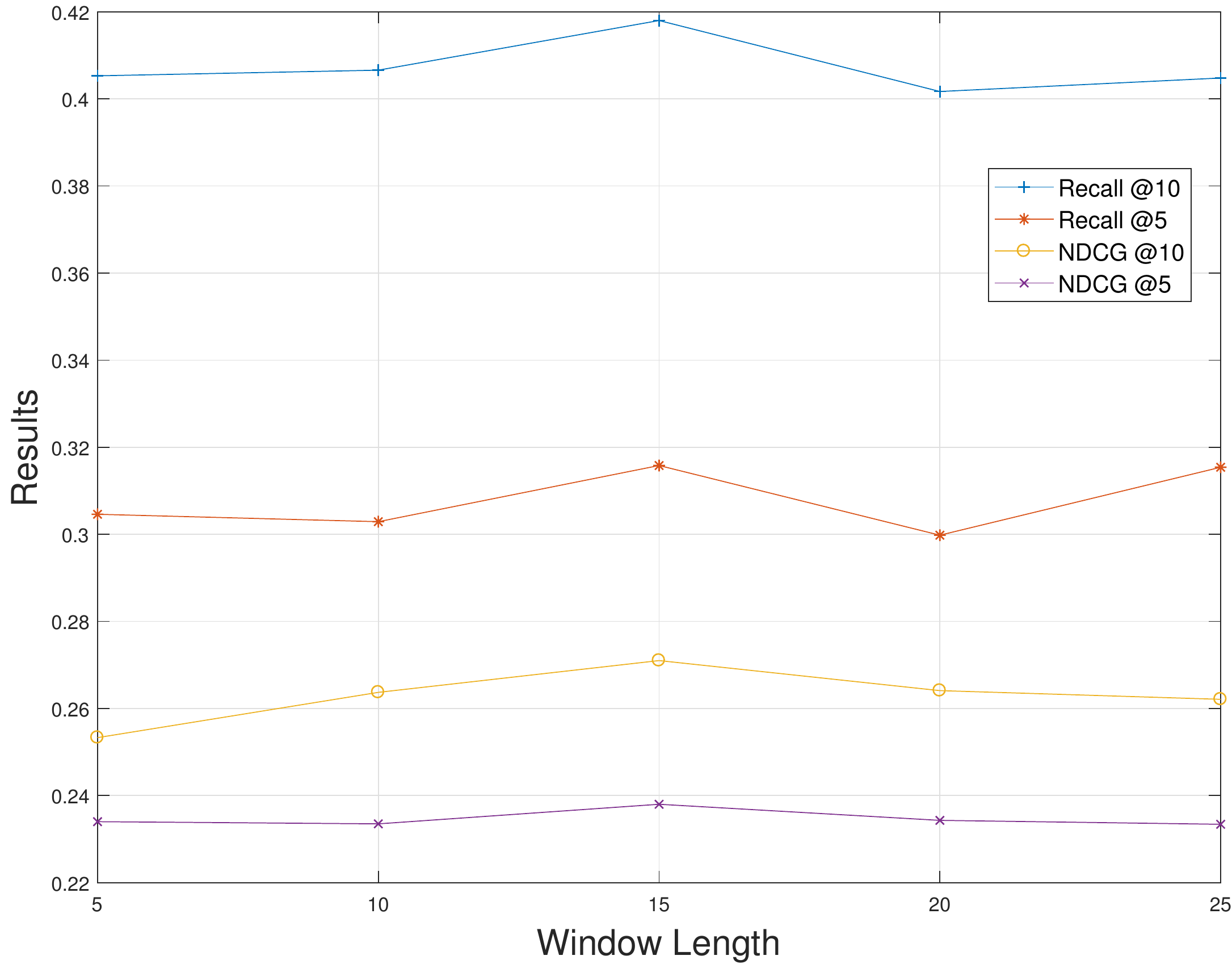}
			\end{minipage}
		}
		\subfigure[Games]
		{
			\begin{minipage}[b]{.315\textwidth}
				\centering
				\includegraphics[width=\textwidth]{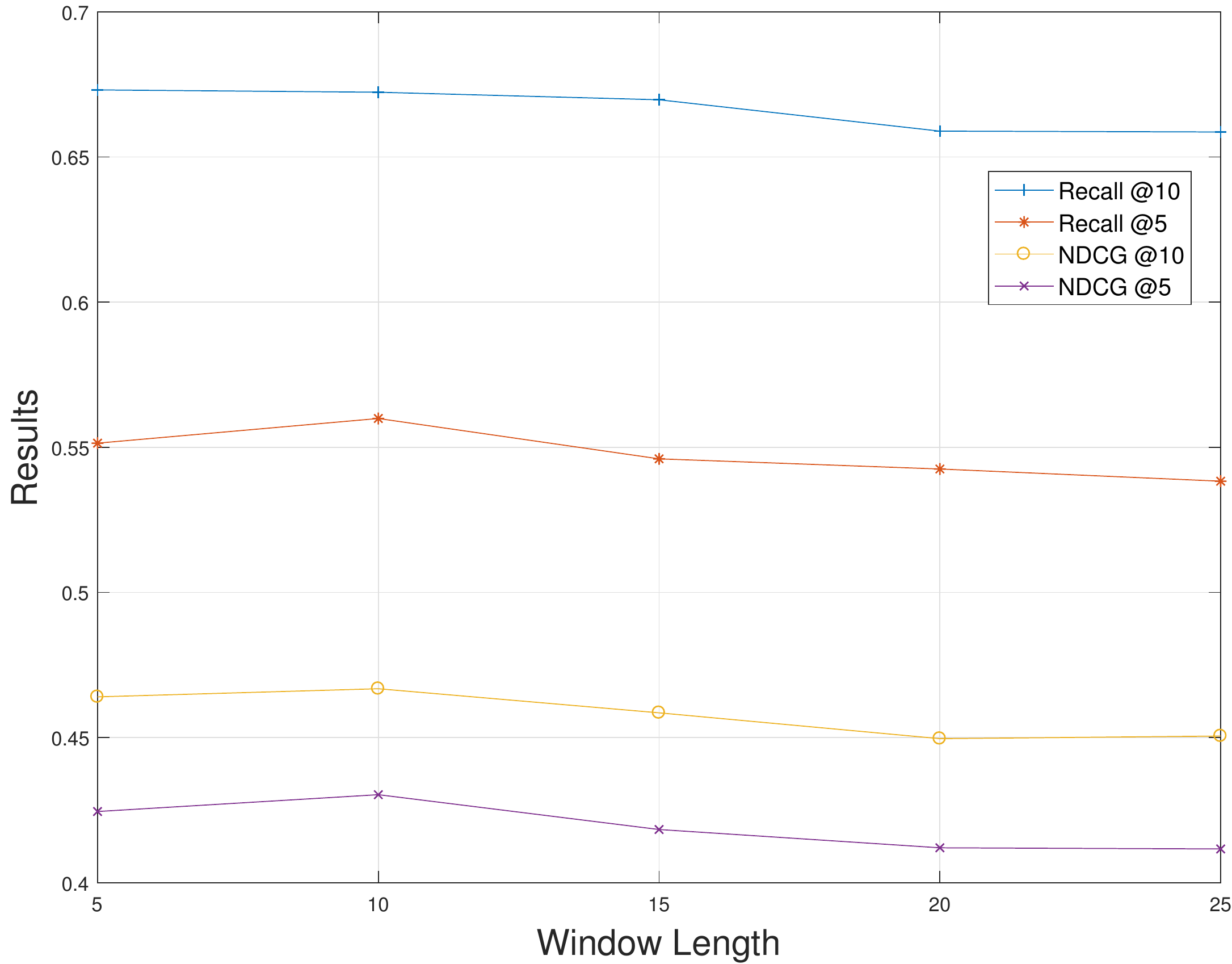}
			\end{minipage}
		}
		\caption{Effect of the length of sliding window}
		\label{window}
	\end{figure*}
	
	\subsection{Hyper-parameter Sensitivity}
	
	The most important hyper-parameters in our model are the number of channels $K$ and the length of sliding window $L$. Specifically, we fix other parameters and adjust the number of one hyperparameter by fixed length. We record the predicting results on the three datasets and draw line charts to show their impacts. We will analyze the figures in this section.
	
	\subsubsection{Impact of the number of channels}  
	We adjust the number of channels from 5 to 35 in steps of 5 and show the results in Figure~\ref{channel}. We can see that the recommendation performance improves as the channel number increases, and tends to remain unchanged after reaching the peak. The MovieLens dataset reaches the peak later than the Amazon dataset. The reason may be that product items do not have as many attributes as the movie items have, and the model does not require too many classifications to achieve the best results.
	
	\subsubsection{Impact of the length of sliding window} 
	We adjust the length of sliding window from 5 to 25 in steps of 5 and show the results in Figure~\ref{window}. We can observe that the influences of the length are quite different from that of channel numbers. On the MovieLens dataset, the performance results become slightly larger as the window length grows, but there is no such trend on the Amazon datasets. Therefore, we can speculate that the choice of window length does not mainly affect the recommended results.

	\section{Conclusion}\label{conclusion}
	
	In this paper, we proposed an edge-enhanced model based on graph neural network to learn sequential representation at both global and local levels. We designed a disentangled learning layer, i.e., the channel-aware mechanism, to distinguish various factors which motivate user intentions. The mechanism divided the information transition model into several channels and aggregated item information through different channels. At the global level, we built a global item-link graph based on training data and update item feature information through neighborhood. At the local level, we apply variational auto-encoder framework to infer user behaviors as distributions, taking advantage of its statistical ability in learning disentangled representation. Then we adopt a sliding window strategy along with the channel-aware mechanism to capture the transition of user intentions through sequences. Experimental results showed that our proposed method achieves better performance than previous works. It is notable that user information is also important for learning disentangled representation. Therefore, we will consider adding user nodes in further studies.
	
	\section*{Acknowledgments}
	This research was partially supported by NSFC (No. 61876117, 61876217, 61872258, 61728205), ESP of the State Key Laboratory of Software Development Environment, and PAPD of Jiangsu Higher Education Institutions.
	
	\balance
	\bibliographystyle{IEEEtran}
	\bibliography{reference}
	
\end{document}